\documentclass[%
 aip,
 amsmath,amssymb,
 reprint,%
]{revtex4-1}

\usepackage{graphicx}% Include figure files
\usepackage{dcolumn}% Align table columns on decimal point
\usepackage{bm}% bold math
\usepackage[utf8]{inputenc}
\usepackage[T1]{fontenc}
\usepackage{mathptmx}
\usepackage{etoolbox}

\makeatletter
\def\@email#1#2{%
 \endgroup
 \patchcmd{\titleblock@produce}
  {\frontmatter@RRAPformat}
  {\frontmatter@RRAPformat{\produce@RRAP{*#1\href{mailto:#2}{#2}}}\frontmatter@RRAPformat}
  {}{}
}%
\makeatother
\begin{document}

\preprint{AIP/123-QED}

\title[Assigning probabilities to non-Lipschitz mechanical systems]{Assigning probabilities to non-Lipschitz mechanical systems}
\author{Danny E.~P.~Vanpoucke}
 \affiliation{UHasselt, Faculty of Sciences, Agoralaan Gebouw D, 3590 Diepenbeek, Belgium
}%
\author{Sylvia Wenmackers}%
 \email{sylvia.wenmackers@kuleuven.be}
\affiliation{KU Leuven, Institute of Philosophy, Centre for Logic and Philosophy of Science (CLPS), Kardinaal Mercierplein 2 - bus 3200, 3000 Leuven, Belgium
}%

\date{\today}

\begin{abstract}
\textbf{This paper was published in \textit{Chaos} 31, 123131 (2021); \href{https://doi.org/10.1063/5.0063388}{https://doi.org/10.1063/5.0063388}.}

\noindent We present a method for assigning probabilities to the solutions of initial value problems that have a Lipschitz singularity.
To illustrate the method, we focus on the following toy example: $\frac{d^2r(t)}{dt^2} = r^a$, $r(t=0) =0$, and $\frac{dr(t)}{dt}\mid_{r(t=0)} =0$, with $a \in ]0,1[$. This example has a physical interpretation as a mass in a uniform gravitational field on a frictionless, rigid dome of a particular shape; the case with $a=1/2$ is known as Norton's dome.
Our approach is based on (1) finite difference equations, which are deterministic; (2) elementary techniques from alpha-theory, a simplified framework for non-standard analysis that allows us to study infinitesimal perturbations; and (3) a uniform prior on the canonical phase space. Our deterministic, hyperfinite grid model allows us to assign probabilities to the solutions of the initial value problem in the original, indeterministic model.
\end{abstract}

\maketitle

\begin{quotation}
Some simple mechanical systems are characterized by indeterministic initial value problems. One such example is `Norton's dome': a mass at rest on a hill of a specific shape in a gravitational field may stay at rest or start sliding off at an arbitrary time. Around a decade ago, this example caught the attention of philosophers of physics, but similar examples were already discussed by mathematicians and physicists in the nineteenth century. We present a numerical simulation study of a class of such mechanical systems. Our approach is to discretize the time variable and to apply results from a branch of mathematics called ‘non-standard analysis’, which allows us to work with infinitesimals in the strict sense (i.e., numbers larger than zero but smaller than $1/n$ for every natural number $n$). This approach also allows us to assign probabilities to the solutions of the indeterministic Cauchy problem, without introducing any non-infinitesimal perturbations (which would defeat the purpose). The methodology may be applicable to study more realistic problems such as turbulent flows, shock waves, and $N$-body collisions.
\end{quotation}

\section{\label{sec:Intro}Introduction}

The nineteenth century mathematician and physicist Poisson was the first to search for a mechanical interpretation of indeterministic Cauchy problems.\cite{Poisson:1806,VanStrien:2014} Later that same century, Boussinesq gave a gravitational interpretation of a broad class of such indeterministic Cauchy problems, by considering a mass placed at rest at the apex of a frictionless surface from a particular family of hill shapes.\cite{Boussinesq:1879a} This work seems to have been largely forgotten, but we want to alert physicists and applied mathematicians to a recent revival of this issue in the philosophical literature: this question was raised again by a contemporary philosopher of science, Norton,\cite{Norton:2003,Norton:2008} who focused on a particularly simple case, now often referred to as \emph{Norton's dome}. Malament generalized Norton's example to a family of problems that we will call \emph{Malament's mounds} (presented in Sec.~\ref{Sec:NortonDome}).\cite{Malament:2008}
These examples involve initial value problems with a differential equation that exhibits a non-Lipschitz singularity.
Such non-Lipschitz Cauchy problems are prevalent in the context of physical applications, such as turbulent flows and associated dispersion,\cite{BenvenisteDrivas:2014} shock waves,\cite{SerreVasseur:2015} and collisions in Newtonian $N$-body problems.\cite{Bakker:2013}
They are also of interest for the foundations of physics, as a case study in determinism and causality, and may be suitable for didactic purposes, to illustrate the role of uniqueness conditions in Newtonian mechanics.
In the context of fluid turbulence, this form of indeterminism has been called `classical spontaneous stochasticity' and it has been proposed that the phenomenon has a quantum-mechanical analog (see Ref.~\citenum{EyinkDrivas:unpublished}, where also the connection to Norton's dome is mentioned).
The main goal of the present paper is to demonstrate a method for assigning probabilities to trajectories that are solutions to non-Lipschitz Cauchy problems.
To demonstrate the method, we focus on the toy problem of Malament's mounds throughout.

Shortly before Norton's work,\cite{Norton:2003,Norton:2008} probabilistic approaches have been applied to non-Lipschitz Cauchy problems;\cite{EVandenEijnden:2000,Falkovich_etal:2001} see more recently also Refs.~\citenum{EVandenEijnden:2003,Mailybaev:2016,Drivas_etal:unpublished}. (We are grateful to an anonymous referee for pointing us to these papers.)
Indeterministic theories can be supplemented by hidden variables to arrive at deterministic theories, which are empirically equivalent and which can be used to assign probabilities to the former (see, \textit{e.g.}, Refs.~\citenum{Werndl:2009} and \citenum{Gisin:2019}). This is the approach we take in this paper, where we let the hidden variables take on infinitesimal values (in the sense of non-standard analysis).
Some physicists presuppose the existence of one unique solution (obtainable, \textit{e.g.}, via physical regularization). However, there are genuine cases of indeterminism, where multiple solutions obtain with various probabilities; therefore, in general, it cannot be taken for granted at the outset that a unique solution dominates. Our method does not start from any \textit{a priori} assumption of a unique solution, although it turns out that probability one should be assigned to a single solution in our case study.

To be clear, our aim is to analyze a class of initial value problems, not any natural phenomena (at least not directly). We start from toy problems, which have inherited the usual idealizations and limits to applicability native to classical physics (\textit{e.g.}, point masses and perfectly frictionless, rigid surfaces). Hence, we are not focusing on when the description is a useful one, and the discussion cannot be settled by direct comparison to experiments.

Our approach relies on an elementary application of non-standard analysis, a branch of mathematics  first developed by \citet{Robinson:1961,Robinson:1966} as an alternative framework for differential and integral calculus The name contrasts with `standard' analysis in terms of the standard real numbers and associated concepts, such as the standard limit (introduced in terms of an epsilon--delta definition) and the derivative and integral defined in terms of this limit. Alternatively, non-standard analysis extends the set of real numbers with infinitely large numbers and infinitesimals, which are closed under the same algebraic rules as the standard reals. Infinitely large numbers are numbers larger than $n$ for all natural numbers $n$; non-zero infinitesimals are their multiplicatory inverse.
The theory uses notions from model theory (a branch of mathematical logic) and is built upon so-called non-standard models of real-closed fields. To keep our paper self-contained yet accessible to non-logicians, we use the framework of alpha-theory, which we introduce in Sec.~\ref{Sec:AlphaTheory}.

Non-standard analysis is often used to obtain results about the real numbers and it contains a theorem that guarantees that the results will agree with those of standard analysis. However, non-standard results can also be studied in their own right, without the end-goal of obtaining a result in terms of standard reals. Moreover, non-standard analysis captures some of the guiding intuitions from the historical development of the calculus, in particular the ideas of Leibniz, some of which were lost during the development of standard analysis in the nineteenth century. (For a brief introduction to the history, see, \textit{e.g.}, Ref.~\citenum{Blaszczyk_etal:2013}.) The idea of infinitesimals in the context of calculus and analysis was long believed to be irreparably confused or intrinsically paradoxical, but this assumption was proven wrong by the work of Robinson in the 1960s\cite{Robinson:1966} and later developments. Finally, non-standard analysis is close to physical praxis and didactics, which in some regard stays close to the Leibnizian appeal to infinitesimals. Indeed, non-standard techniques have been applied to physics in a variety of applications, including Brownian motion, perturbation theory for differential equations, \textit{etc}.\cite{Albeverio_etal:1986} Many of these applications involve a discrete model with infinitesimal steps of quantities that are taken to be continuous in the standard model. In particular, we will consider difference equations on discrete grids with infinitesimal time steps.

Since our method relies on non-standard analysis, it can yield genuinely new results only as long as the results are presented in terms of hyperreal numbers. Once the results are interpreted in the context of standard real numbers, it cannot yield anything that cannot be obtained via methods of standard analysis. However, even in that case, it may still be relevant, since the non-standard approach may be shorter, easier to obtain or more instructive than the standard one. Here, we argue that the non-standard approach suggests a way of assigning probabilities to the standard solutions of indeterministic initial value problems. Perhaps this will inspire future work that achieves similar results without having to introduce non-standard methods.

%Structure of the paper
Our paper is structured as follows. Section~\ref{Sec:NortonDome} reviews the shape, initial value problem, and standard solutions for Malament's mounds.
In Sec.~\ref{Sec:ResearchQ}, we refine our research questions and specify our working hypothesis.
In Sec.~\ref{Sec:AlphaTheory}, we introduce concepts from alpha-theory, a simplified approach to non-standard analysis.
In Sec.~\ref{Sec:HyperfiniteDome}, we apply this to build an alternative model for Malament's mounds, in which we can consider infinitesimal perturbations. This approach allows us to assign probabilities to the standard solutions in Sec.~\ref{Sec:Probability}.
We offer some discussion and review our main conclusions in Sec.~\ref{Sec:Disc}.

\section{Malament's Mounds\label{Sec:NortonDome}}
\begin{figure}[!htb]
\centering
  \includegraphics[width=0.48\textwidth]{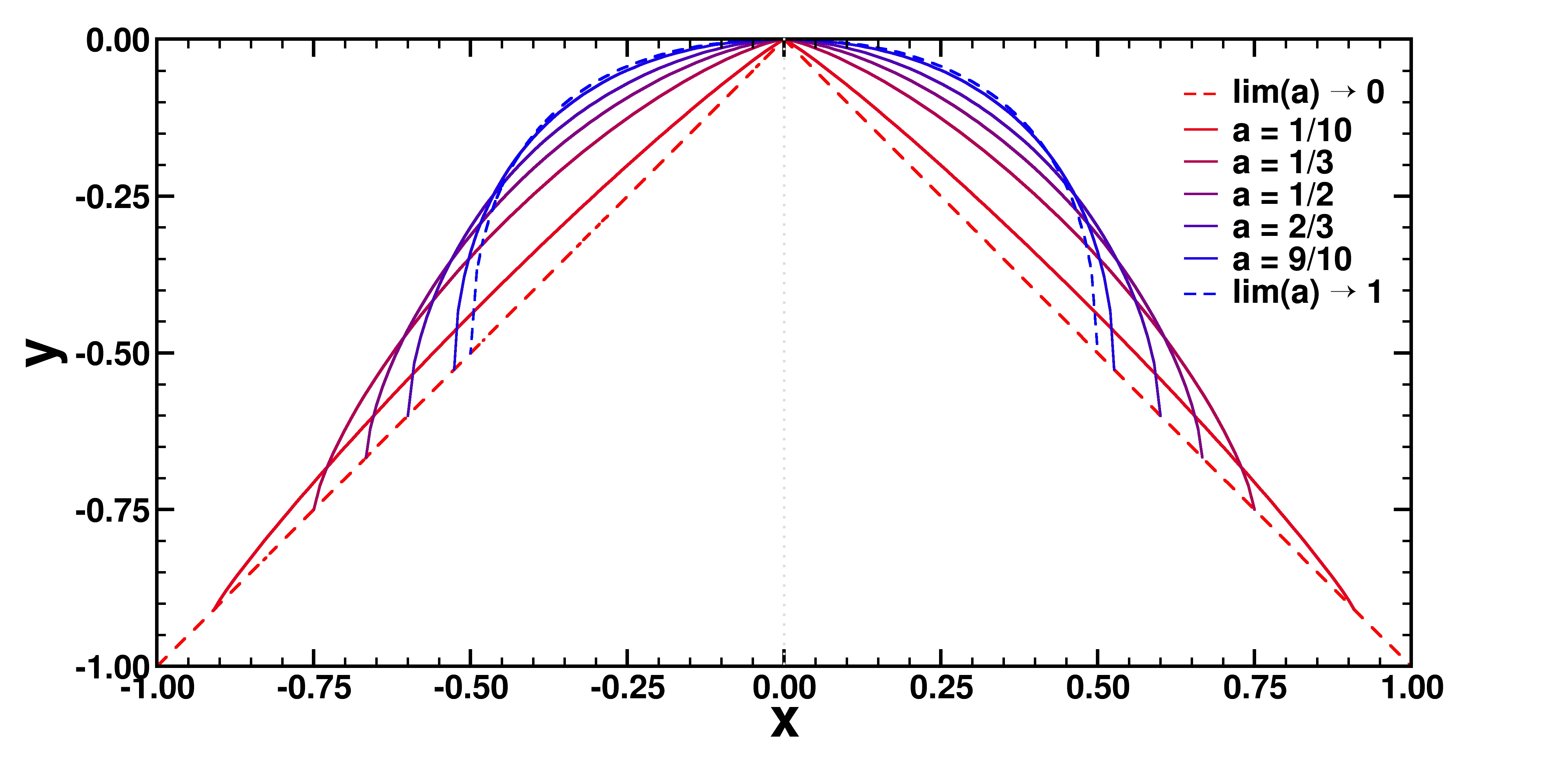}\\
  \caption{Cross section of five of Malament's mounds: $a=1/10$, $a=1/3$, $a=1/2$ (Norton's dome), $a=2/3$, and $a=9/10$. The limiting cases with $a \rightarrow 1$ (a mound of height 1/2) and $a \rightarrow 0$ (triangle of height 1) are shown as dashed curves. Observe that the base points of Malament's mounds follow this limiting triangle.}\label{Fig:Fig3}
\end{figure}

Norton's problem represents a mass placed with zero velocity at the apex of a particular dome in a uniform gravitational field. The shape of the dome is chosen such that Newton's second law applied to the mass takes on a particularly simple form, as we will see below.
Malament generalized Norton's dome to the following family of shapes,\cite{Malament:2008} which yields a family of indeterministic Cauchy problems:

\begin{equation*}
y(x) = - \frac{\left(1 - \left(1 - \left(1+a \right) |x| \right)^\frac{1}{1+a}\right)^{1+a}}{1+a},
\end{equation*}
where $a$ is any real number in $]0,1[$, $x$ is the horizontal axis (orthogonal to the gravitational field), $y$ is the vertical height (anti-parallel to the gravitational field), and the apex is at the point $(0,0)$.
See Fig.~\ref{Fig:Fig3} for five examples of hill shapes.
Observe that the above expression becomes undefined for $x$-values larger than $1/(1+a)$; therefore, the mounds have a maximal height and unilateral width of $1/(1+a)$.

Define $r \geq 0$ as the arc distance measured along the dome from the apex. Then, we find:
\begin{equation*}
r(x) = 1 - \left(1 - (1+a)|x| \right)^{\frac{1}{(1+a)}}.
\end{equation*}

Expressing $y$ as a function of the arc length $r$ measured from the apex yields:
\begin{equation*}
y(r) = - \frac{r^{1+a}}{(1+a)}.
\end{equation*}

We assume that the gravitational field is constant with $g=1$ and that a unit mass moves on a frictionless hill of the specified family.
Then, Newton's second law yields a second-order non-linear ordinary differential equation (ODE) involving a non-Lipschitz continuous function.
For each choice of $a \in ]0,1[$, the Cauchy problem for the corresponding Malament mound is given by:

\begin{equation}
\left\{
\begin{array}{cl}
\frac{d^2r(t)}{dt^2} &= r^a \\
r(t=0) &=0 \\
\frac{dr(t)}{dt}\mid_{r(t=0)} &=0.
\end{array}
\right.\label{Eq:MalamentMounds}
\end{equation}

\begin{figure}[!htb]
\centering
  \includegraphics[width=0.4\textwidth]{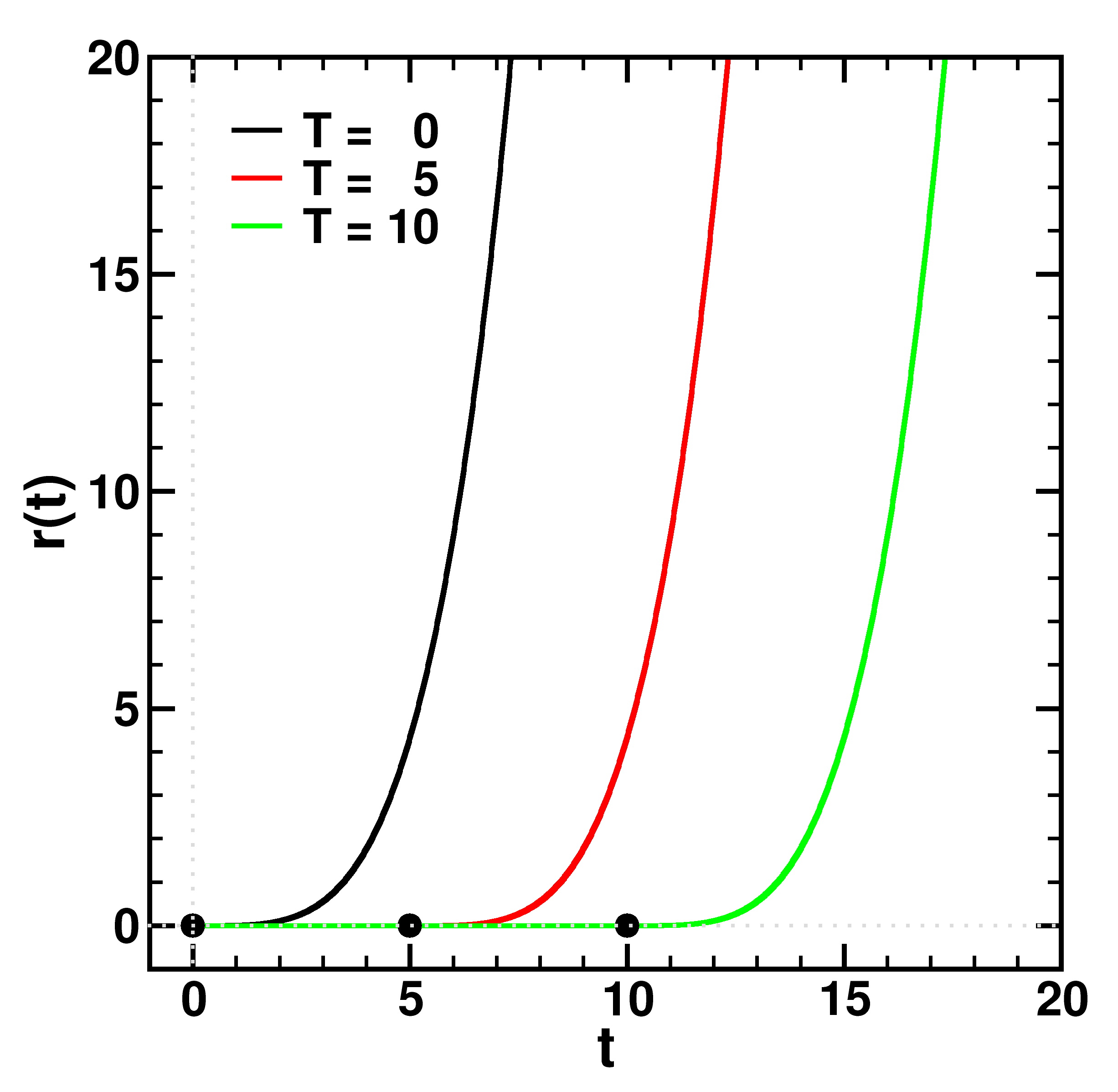}\\
  \caption{Three regular solutions to Norton's dome ($a=\frac 12$) with $T$ equal to 0 (black), 5 (red), and 10 (green).}\label{Fig:Fig2}
\end{figure}

This problem is the second-order analog of a textbook example commonly used to illustrate a failure of Lipschitz continuity.
As is well known, the solution of such problems is non-unique.
Besides the trivial, singular solution, $r(t)=0$, there is a one-parameter family of regular solutions (see, \textit{e.g.}, Theorem~2 in Ref.~\citenum{Derks:2009}, due to Kneser\cite{Kneser:1923}), which can be represented geometrically as a \emph{Peano broom} (see Fig.~\ref{Fig:Fig2}),

\begin{equation}
r(t)=\left\{
\begin{array}{cl}
0 & \textrm{if\ } t \leq T \\
\left( \frac{(1-a)^2}{2(1+a)} \right)^{\frac{1}{1-a}}(t-T)^{\frac{2}{1-a}} & \textrm{if\ } t \geq T ,%
\end{array}
\right.\label{Eq:MalamentSolution}
\end{equation}
\noindent where $T$ is a positive real number, which can be interpreted as the time at which the mass starts sliding off the hill.
The solution can be verified by substitution into (\ref{Eq:MalamentMounds}).

In the three-dimensional case, there is an additional continuum of possibilities regarding the direction of descent. Throughout this paper, we limit ourselves to the two-dimensional case (as depicted in Fig.~\ref{Fig:Fig3}) such that this indeterminacy is reduced to two possible directions.

\section{Research Questions and Working Hypothesis\label{Sec:ResearchQ}}
Faced with indeterminism due to a lack of Lipschitz continuity, some authors search for arguments that single out a unique solution, \textit{e.g.}, by regularization (smoothing the system close to the singularity) or adding physical principles or heuristics not encoded by the Cauchy problem itself. The assumption that there is one correct solution (motivated by additional physical constraints besides the mathematical equation) is widely---though perhaps not univocally---held in the field of fluid dynamics. For instance, the authors of Ref.~\citenum{DrivasMailybaev:2021} aim to regulate the solutions of Cauchy problems with non-Lipschitz indeterminism to select a unique global solution.

Our current approach is slightly different: lacking a unique solution, we look for a probabilistic description for the trajectory of the mass. This approach may be alien to Newtonian physics (the context in which the problems of Sec.~\ref{Sec:NortonDome} arose), but it is accepted in classical physics more generally (i.e., in statistical physics). Hence, we start with the following research question:
\begin{itemize}
  \item Given that there are multiple solutions to Cauchy problem (\ref{Eq:MalamentMounds}), is there a well-supported way to assign probabilities to them?
\end{itemize}
This research question leads us to two more specific questions:
\begin{itemize}
  \item Can we assign relative probabilities to the singular solution vs the family of regular solutions?
  \item Can we assign relative probabilities to the various regular solutions (regarding $T$ and the direction of movement)?
\end{itemize}

Since our questions are aimed at finding probabilities, the usual approach of physical regularization is of no avail here (but see Sec.~\ref{Sec:Hydrodyn}). Alternatively, we aim to represent the trajectories on Malament's mounds using a discrete, deterministic model from non-standard analysis that can help us to measure the sought probabilities.

\section{Alpha-theory, $\alpha$-limits, and hyperfinite grid differential equations\label{Sec:AlphaTheory}}

In this section, we first introduce a simplified framework for non-standard analysis: alpha-theory, which was developed by Benci and Di Nasso.\cite{BenciDiNasso:2019} We then show how their notion of hyperfinite grid differential equations is used to find all solutions to indeterministic Cauchy problems and how to assign probabilities to them.

Alpha-theory defines a new, ideal number, $\alpha$, which can be thought of as an infinite number (larger than every natural number) that captures the rate of divergence of the linear sequence $1,2,3, \ldots$. The theory also defines a new type of limit, the $\alpha$-limit, which can be thought of as the value a sequence would take if it was extended to position $\alpha$. This limit is not to be confused with the standard limit operation: except for the special case of constant sequences of real numbers, they do not agree. (See the end of Sec.~\ref{Sec:alphalimits} for some examples.) Moreover, the $\alpha$-limit is broader in scope: it does not only apply to sequences of real numbers, but to sequences of any type of objects, including sets and hyperreal numbers.
We choose alpha-theory because it suffices to introduce the notion of hyperfinite grid functions and associated differential equations.\cite{BenciDiNasso:2019,Benci_etal:2010} These hyperfinite grid differential equations behave much like finite difference equations, except that the number of steps is infinite and each step is infinitesimal.

\subsection{$\alpha$-limits\label{Sec:alphalimits}}
Alpha-theory assumes most of standard mathematics (more specifically, it is based on Zermelo--Fraenkel set theory with the axiom of choice, but without the axiom of regularity) to which it adds six new axioms (see pp.~77--78 in Ref.~\citenum{BenciDiNasso:2019}), which we reproduce here for self-containedness:
\begin{enumerate}
  \item Every sequence $A(n)$ has a unique \emph{$\alpha$-limit}, denoted by $\lim\limits_{n\uparrow \alpha} A(n)$. If $A(n)$ is a sequence of atoms (i.e., primitive elements that are not sets), then $\lim\limits_{n\uparrow \alpha} A(n)$ is an atom, too.
  \item If $R_r(n)=r$ is a constant sequence with the real number $r$ as its value, then $\lim\limits_{n\uparrow \alpha} R_r(n) = r$.
  \item The $\alpha$-limit of the identity sequence $I(n) = n$ is a new number $\alpha$ such that $\lim\limits_{n\uparrow \alpha} I(n) = \alpha \notin \mathbb{N}$.
  \item The set of all $\alpha$-limits of real sequences is called the set of \emph{hyperreals}, denoted by 
  $${^* \mathbb{R}} = \left\{ \lim\limits_{n\uparrow \alpha} R(n) \middle| R: \mathbb{N} \rightarrow \mathbb{R} \right\}.$$
  $\langle {^* \mathbb{R}}; +; \times; 0; 1 \rangle$ forms a field, called the hyperreal field, where $\lim\limits_{n\uparrow \alpha} R_1(n) + \lim\limits_{n\uparrow \alpha} R_2(n) = \lim\limits_{n\uparrow \alpha} (R_1(n)+R_2(n))$ and $\lim\limits_{n\uparrow \alpha} R_1(n) \times \lim\limits_{n\uparrow \alpha} R_2(n) = \lim\limits_{n\uparrow \alpha} (R_1(n) \times R_2(n))$.
  \item The $\alpha$-limit of the constant sequence with the value equal to the empty set, $S_\emptyset(n) = \emptyset$, equals the empty set; i.e., $\lim\limits_{n\uparrow \alpha} S_\emptyset(n) = \emptyset$. The $\alpha$-limit of a sequence of non-empty sets $S(n)$ is $\lim\limits_{n\uparrow \alpha} S(n) = \left\{ \lim\limits_{n\uparrow \alpha} R(n) \middle| \forall n R(n) \in S(n) \right\}$.
  \item If $A(n)$ and $B(n)$ are two sequences such that $\lim\limits_{n\uparrow \alpha} A(n) = \lim\limits_{n\uparrow \alpha} B(n)$ and $f$ is a function such that the compositions $f \circ A$ and $f \circ B$ make sense, then $\lim\limits_{n\uparrow \alpha} f(A(n)) = \lim\limits_{n\uparrow \alpha} f(B(n))$.
\end{enumerate}

Observe that axiom 4 ensures that the usual algebraic operations on the reals are well-defined for the hyperreals, too.

Some additional definitions are helpful.\cite{BenciDiNasso:2019}
\begin{enumerate}
  \item A hyperreal number $\rho$ is called \emph{infinite} if there exists no standard real $r$ such that $r>|\rho|$; otherwise, the hyperreal is called \emph{finite}. A hyperreal number $\epsilon$ is called \emph{infinitesimal} if there exists an infinite hyperreal $\rho$ such that $\epsilon=1/\rho$.
  \item For any object $A$, its \emph{hyper-image} (or $^*$-transform) is defined as: $${^* A} = \left\{ \lim\limits_{n\uparrow \alpha} S(n) \middle| S : \mathbb{N} \rightarrow A \right\}.$$ This extends the meaning of the `hyper-' (or ${^*}$-) prefix already introduced in axiom 4 for the set of hyperreals to all sets and other objects.
  \item A set $T \subset {^* S}$ is \emph{hyperfinite} if there exists a sequence of finite sets $S_n \subset S$ and a sequence $U(n) : \mathbb{N} \rightarrow S_n$ such that $ T = \left\{ \lim\limits_{n\uparrow \alpha} U(n) \right\}.$
  \item The \emph{hyperfinite sum} of a hyperfinite set of hyperreal numbers, $T = \lim\limits_{n\uparrow \alpha} U(n)$, is defined as
  $$\sum_{\rho \in T} \rho = \lim\limits_{n\uparrow \alpha} \sum_{r \in U(n)} r.$$
%  \item An object is called `internal' if it is the $\alpha$-limit of some sequence. (Benci and Di Nasso, 2019 also prove that this definition agrees with the notion of internal objects in non-standard analysis.)
\end{enumerate}

Finally, we include three theorems that will be useful for our problem. (For proofs, see p.~9, pp.~288--289, and p.~92 of Ref.~\citenum{BenciDiNasso:2019}, respectively.)
\begin{enumerate}
  \item Every finite hyperreal number $\rho$ is infinitesimally close to a unique real number $r$, called its \emph{standard part}: $r = st(\rho)$.
  In particular, if $\epsilon$ is infinitesimal, then its standard part equals zero: $st(\epsilon)=0$. Therefore, taking the standard part can be thought of as `rounding off' the infinitesimal part.
  \item $\alpha$ can consistently be assumed to be a multiple of each natural number and each natural power.
  \item For any function $f : D \rightarrow C$, its hyper-image is a function ${^* f} = {^*D} \rightarrow {^*C}$ such that for every sequence $R : \mathbb{N} \rightarrow D$, it holds that ${^* f}\left( \lim\limits_{n\uparrow \alpha} R(n)\right) = \lim\limits_{n\uparrow \alpha} {^* (f \circ R)} (n)$.
\end{enumerate}

It is instructive to compare the $\alpha$-limit of a few sequences.
The constant sequence $R_0(n) = 0$, the linearly decreasing sequence $R_l(n)=\frac 1n$, and the quadratically decreasing sequence $R_q(n)= \frac{1}{n^2}$ all have the same standard limit, 0.
However, they have different $\alpha$-limits: $\lim\limits_{n\uparrow \alpha} R_0(n) = 0$, $\lim\limits_{n\uparrow \alpha} R_l(n) = \frac{1}{\alpha}$, and $\lim\limits_{n\uparrow \alpha} R_q(n) = \frac{1}{\alpha^2}$. The latter two are infinitesimals, with $\frac{1}{\alpha} > \frac{1}{\alpha^2}$.
Still, the standard part of both these hyperreals is zero.
Therefore, one way to interpret the hyperreals is that---compared to real numbers---they retain information about the asymptotic behavior of the sequences by which they were constructed.
Likewise, infinite hyperreals contain information about the rate of divergence of the sequence by which they are obtained. For example, the $\alpha$-limit of the sequence $n^2$ is $\alpha^2$, the square of that of $I(n)=n$ which is $\alpha$. This is similar to Landau's symbols (small-$o$ and big-$O$ notation), but the hyperreal field provides a richer algebraic structure.

\subsection{Functions, derivatives, and differential equations on hyperfinite grids}
To construct a hyperfinite grid, which we will use to model time, we need to consider the $\alpha$-limit of a sequence of sets. A first example of the $\alpha$-limit applied to a sequence of sets (proven on p.~81 of Ref.~\citenum{BenciDiNasso:2019}) is $\lim\limits_{n\uparrow \alpha} \{1, \ldots, n\} = \{1, \ldots, \alpha \}.$ This set is hyperfinite, but not finite.\cite{BenciDiNasso:2019}

A sequence of sets of interest to our problem at hand is
$$M(n) = \left\{ 0,\frac{1}{n}, \ldots, n-\frac{1}{n}, n \right\}.$$
We may think of each $M(n)$ as set of discrete moments (a \emph{finite grid}). As $n$ increases, the sets contain more moments per unit of time and span a longer time. The $\alpha$-limit of this sequence is a hyperfinite set, which we call a \emph{hyperfinite grid},
$$\mathbb{M} = \lim\limits_{n\uparrow \alpha} M(n) = \left\{ 0, \frac{1}{\alpha}, \ldots, \alpha-\frac{1}{\alpha}, \alpha \right\}.$$
$\mathbb{M}$ contains infinitely many moments per unit of time and infinitely many such units. The infinitesimal time step between two consecutive elements of $\mathbb{M}$ has length $1/\alpha$. Intuitively, then, the discrete set $\mathbb{M}$ can be used to represent the positive direction of time, just like the continuous set $\mathbb{R}^+$ often plays this role.
If we assume $\alpha$ to be a multiple of each natural number and each natural power, then $\mathbb{M} \cap \mathbb{R} = \mathbb{Q}^+$.

We will call a function $R: \mathbb{M} \rightarrow {^* \mathbb{R}}$, which is the $\alpha$-limit of a sequence of functions $R_n : M(n) \rightarrow \mathbb{R}$, a \emph{grid function}.

For our problem, we also need to define the first and second-order grid derivative of a grid function, $R$ (see pp.~160--161 in Ref.~\citenum{BenciDiNasso:2019}; the definitions are analogous to Euler's method).
First, consider a family of functions $R_n : M(n) \rightarrow \mathbb{R}$, with the grid function $R$ as their $\alpha$-limit. 
The \emph{right-hand grid derivative} $D^+ R$ is then defined as the $\alpha$-limit of the sequence of first-order finite differences on the $R_n$, which, at position $m \in M_n$, equal $n \left( R_n(m+1) - R_n(m)\right)$. The factor $n$ comes from division by the discrete time step, $1/n$. Hence, in the $\alpha$-limit,
$$D^+ R(m) = \alpha \left( R(m+1) - R(m) \right) ,$$
for all $m \in \mathbb{M} \setminus \{ \alpha \}$.
Likewise, the \emph{second-order grid derivative}, $\Delta R$, is defined as the $\alpha$-limit of the sequence second-order finite differences on the $R_n$, which, at position $m \in M_n$, equal $n^2 \left( R_n(m+1) - 2 R_n(m) + R_n(m-1) \right)$. Hence,
$$\Delta R(m) = \alpha^2 \left( R(m+1) - 2 R(m) + R(m-1) \right) ,$$
where $m \in \mathbb{M} \setminus \{ 0, \alpha \}$.

With each standard function $f : \mathbb{R}^+ \rightarrow \mathbb{R}$ we may associate a grid function $R_f$, by restricting its hyper-image ${^* f}$ to $\mathbb{M}$ (cf.\ p.~160 in Ref.~\citenum{BenciDiNasso:2019}).
It can be proven that where $\frac{df(t)}{dt}$ and $\frac{d^2f(t)}{dt^2}$ are defined, $\forall t \in \mathbb{Q}^+$,
$$st(D^+ R_f(m)) = \frac{df(t)}{dt}$$
$$st(\Delta R_f(m)) = \frac{d^2f(t)}{dt^2},$$
with $m$ such that $t = st(m/\alpha)$. (Proof is given on p.~161 of Ref.~\citenum{BenciDiNasso:2019}.)

With a given standard Cauchy problem, we can now associate a \emph{hyperfinite grid differential equation} with two initial conditions on the grid function. This `association' is one-to-many because the standard initial conditions can be precisified in infinitely many ways. In particular, there are infinitely many hyperreals $m_0$, $R(0)$, and $R(1)$ such that $t_{0} = st(m_0/\alpha)$, $r(t_0)=st(R(0))$, and $\frac{dr(t)}{dt}\mid_{r(t=0)} = st(\alpha (R(1)-R(0)))$.
Each choice of $m_0$, $R(0)$, and $R(1)$ leads to a different solution in terms of a grid function.
When the standard Cauchy problem has a unique solution, then all these grid functions are infinitesimally close; hence they have the same standard part. It can be shown that the standard part of the grid solution equals the solution to the standard Cauchy problem.
However, when the standard Cauchy problem fails uniqueness, the associated grid functions contain pairs that differ by more than an infinitesimal from each other; hence they have different standard parts.
The approach we sketched here is based on the idea of the non-standard proof for the standard Peano theorem (see, \textit{e.g.}, pp.~165--167 in Ref.~\citenum{BenciDiNasso:2019}, p.~32 in Ref.~\citenum{Albeverio_etal:1986}, and Ref.~\citenum{BirkelandNormann:1980}), which---unlike the standard proof---shows us how to construct all these solutions. We apply this to our case in Sec.~\ref{Sec:HyperfiniteDome}.

Moreover, we can use this approach to associate a probability measure to the different standard solutions.
Recall that the different infinitesimal precisifications can be obtained as $\alpha$-limits of different converging sequences, which also correspond to all the different ways one could take the standard limit (and which may lead to different standard outcomes). Rather than arguing for one limit process as the correct one, we take a measure over all possible ways of converging.
This idea is not entirely novel, although its application to non-Lipschitz continuity is: a similar approach was proposed in the context of stochastic differential equations.\cite{Benci_etal:2010}
%Concerning a noise term $\zeta$, they wrote: ``if $\zeta$ is regarded as a random variable, the probability on the sample space $\mathcal{R}$ can be defined in a naive way, namely every noise has the same probability. This is the basic idea of the Loeb measure''.
%On p.24, they write: ``probability can be introduced in a very elementary way. We may think of the stochastic class $\mathcal{R}$$ as a sample space. The events are the hyperfinite sets $E \subset \mathcal{P}(\mathcal{R}){^*}$$ and the probability P of an event is given by $P(E) = |E|/|\mathcal{R}|$.'' (Notice that they use the suffix notation for hyper-images.)
We will explain this in Sec.~\ref{Sec:Probability}.

\section{Hyperfinite grid differential equation with initial conditions for Malament's mounds\label{Sec:HyperfiniteDome}}
Let us now apply the approach outlined in Sec.~\ref{Sec:AlphaTheory} to our standard Cauchy problem.
We will construct all solutions to this problem, i.e., functions $r(t): \mathbb{R}^+ \rightarrow \mathbb{R}^+$, such that the set of equations (\ref{Eq:MalamentMounds}) hold, using a hyperfinite grid equation.

Each hyperfinite grid Cauchy problem associated with our standard problem looks as follows
\begin{equation}
\left\{
\begin{array}{cl}
\Delta R(m) &= R^a \\
D^+ R(m_0) &= V_0 \\
R(m_0) &= R_0  ,%
\end{array}
\right.\label{Eq:HyperfiniteGridCauchy}
\end{equation}
where $m_0$, $R_0$, and $V_0$ are infinitesimals.
A solution to this problem is a grid function $R: \mathbb{M} \rightarrow {^*\mathbb{R}}$ such that (\ref{Eq:HyperfiniteGridCauchy}) holds.

In order to find these solutions, we construct a sequence of finite difference equations on the set of moments $M_n$ with associated initial conditions.
The finite difference equation on $M_n$ associated with our ODE is, for all $m \in \{1,...,n-1\}$: 

$$n^2 \left( R_n(m+1) - 2 R_n(m) + R_n(m-1) \right) = R_n(m)^a .$$

To make explicit that the solution can be obtained by iteration, we rewrite this as a recurrence relation and replace $m$ by $m-1$ in all terms (so now $m \in \{2, \ldots, n\}$).
Adding the initial conditions, we obtain the following sequence of discrete initial value problems:
\begin{equation}
\left\{
\begin{array}{cl}
R_n(m) &= \frac{1}{n^2} R_n(m-1)^a + 2 R_n(m-1) - R_n(m-2) \\
R_n(1) &= R_{n,1} \\
R_n(0) &= R_{n,0}  ,%
\end{array}
\right.\label{Eq:FiniteGridCauchy}
\end{equation}
where $R_{n,0}$ and $R_{n,1}$ are real-valued sequences that converge to 0 as $n$ goes to infinity; therefore, their $\alpha$-limits are infinitesimal.
Notice that we can continue $R$ for arbitrarily large values, i.e., beyond the maximal height of the mounds. However, we are interested only in the region around the apex.

In general, one should also consider the sequences $m_{n,0} \in M_n^\mathbb{N}$ that converge to 0 as the initial moment (instead of fixing this at $m=0$ for all $n$), but for autonomous equations such as ours this does not lead to more than an infinitesimal difference in the $\alpha$-limit; therefore, it does not impact the standard solution.

For each choice of the initial conditions, $R_{n,0}$ and $R_{n,1}$, the solution is a unique sequence $ R_n(m): M_n \rightarrow \mathbb{R}^+$ that can be obtained recursively. When $R_{n,0} = R_{n,1} = 0$, we obtain the constant sequence $R_n(m)=0$ for all $n$, which corresponds to $R(m)=0$ in the $\alpha$-limit and leads to the singular standard solution $r(t)=0$ after taking the standard part.

What if $R_{n,0}$ or $R_{n,1}$ or both are non-zero? Unfortunately, there is no analytic solution known for the non-linear difference equation of second order in (\ref{Eq:FiniteGridCauchy}); therefore, we will have to examine its behavior numerically.
However, notice that we can read off the scaling behavior of the non-zero solutions from the difference equation directly, by noticing that terms of the form $\frac{1}{n^2} R_n^a$ are added to terms of the form $R_n$. Hence, by the principle of dimensional homogeneity, expressing $R_n$ as multiples of $n^{-\frac{2}{(1-a)}}$ is helpful because this factors out in all terms. This scale factor will also play a crucial role in taking the $\alpha$-limit.

\subsection{Numerical study of the family of finite difference equations\label{Sec:Euler}}

Numerically, we find that for any value of $a$, $R_{n,0}$, and $R_{n,1}$ we can fit the numerical solutions of (\ref{Eq:FiniteGridCauchy}) to
\begin{equation*}
F^a_n(m)= \left( \frac{(1-a)^2}{2(1+a)} \right)^{\frac{1}{1-a}} \left( \frac{m - M}{n} \right)^{\frac{2}{1-a}},
\end{equation*}
for some real value of $M$ and this fit improves as $m$ increases. Therefore, from now on we will only consider the final value of $m$, which equals $n^2$. Hence, we can estimate $T$ as $M/n$, where

\begin{equation}
M = n^2 - n \left( \left( \frac{2(1+a)}{(1-a)^2} \right)^{\frac{1}{1-a}} R_n(n^2)\right)^{\frac{-2}{1-a}}. \label{Mfit}
\end{equation}

\subsubsection{Results for $V_{n,0}=0$\label{Sec:resultsVn0is0}}
First, we study the special case where $R_{n,0}=R_{n,1}$; hence, the initial grid velocity is zero.

For example, if we consider $a=1/2$ and $R_{n,0}=R_{n,1}=n^{-\frac{2}{(1-a)}}$ (the scale factor), the fitted $M$ is about $-1.947$. Therefore, for instance when $n=100$, $T=-0.019~47$.
This value is negative, meaning that, at large $n$, the fitted curve looks as though the mass left the top before $t=0$. This is exactly as you would expect for starting positions that are relatively far from the apex. In terms of direction, the mass will slide off exactly in the direction of the initial displacement.

\begin{figure}
\centering
    \includegraphics[width=0.48\textwidth]{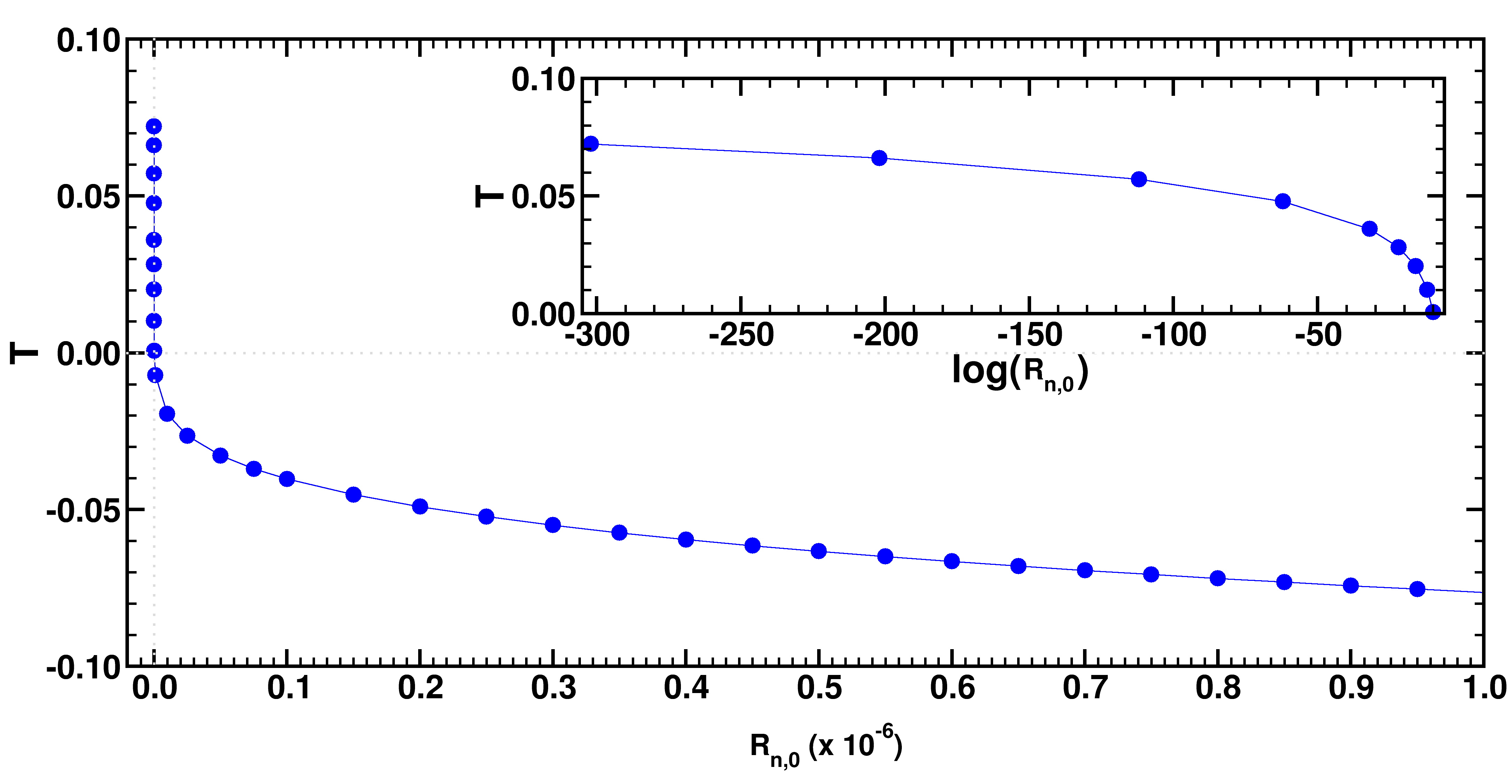}
  \caption{Illustration of $T$ for the special case where $R_{n,0}=R_{n,1}$. These $T$-values were computed as $M/n$ via (\ref{Mfit}) with $a=1/2$ and $n=100$. The inset gives the positive $T$-values on a logarithmic $R_{n,0}$-axis.}\label{Fig:Fig4}
\end{figure}

For values of $R_{n,0}=R_{n,1}$ that are smaller, we find other fitted solutions with a less negative or even positive $T$-value. Positive $T$-values correspond to fitted curves that, at large $n$, look as though the mass left the top later than $t=0$.
Therefore, the smaller the perturbation, the more apparent delay, as expected. Moreover, the relation between the size of the perturbation and the $T$-value of the corresponding solution is highly non-linear. This is illustrated, for the case $R_{n,0}=R_{n,1}$, in Fig.~\ref{Fig:Fig4}.

\begin{figure}
\centering
    \includegraphics[width=0.48\textwidth]{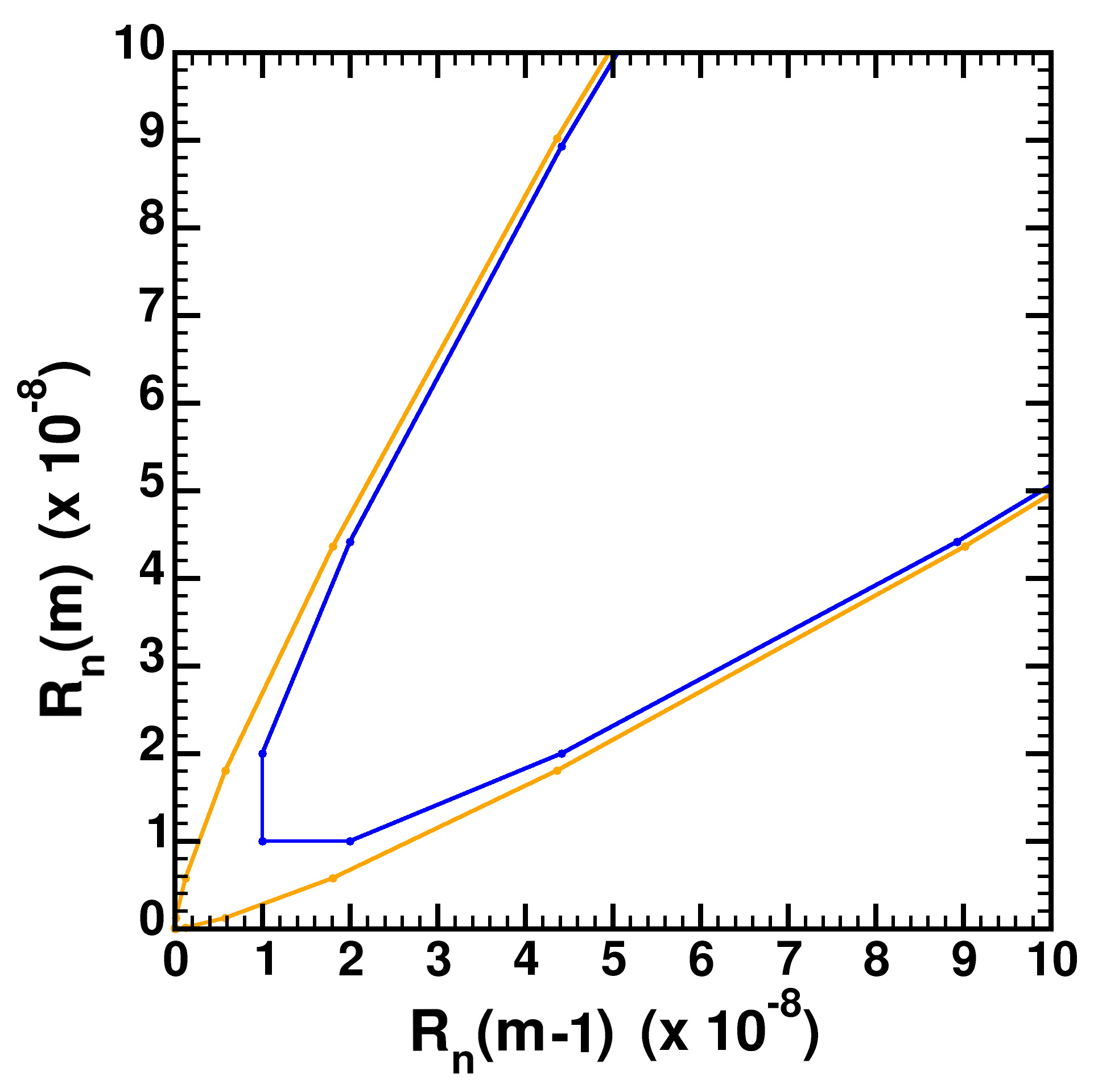}
  \caption{Trajectories are pairs of $(R_n(m-1),R_n(m))$, where $R_n(m) = \frac{1}{n^2} R_n(m-1)^a + 2 R_n(m-1) - R_n(m-2)$ with $a=1/2$ and $n=100$. The blue trajectory passes through the point $R_n(m-1)=R_n(m)=10^{-8}$: its minimal distance to the apex as visible here. The orange trajectory passes through the point $R_n(m-1)=R_n(m)=10^{-16}$, much closer to the origin.}\label{Fig:Fig5}
\end{figure}

\begin{figure}
\centering
    \includegraphics[width=0.48\textwidth]{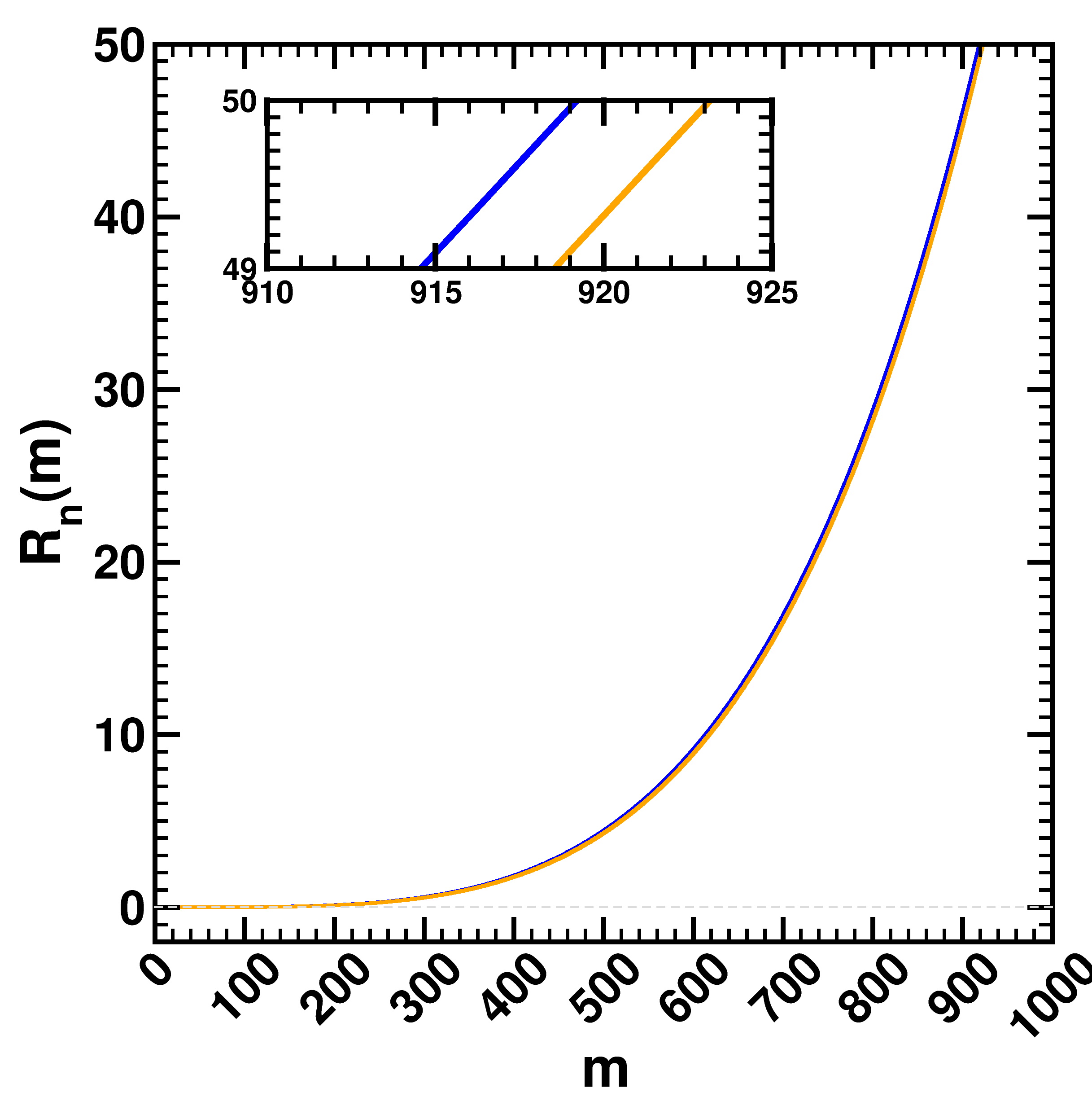}
  \caption{Values for $R_n(m)$ as a function of $m\ (=n t)$ for large $m$, where $R_n(m) = \frac{1}{n^2} R_n(m-1)^a + 2 R_n(m-1) - R_n(m-2)$ with $a=1/2$ and $n=100$. The blue trajectory passes through the point $R_{n,0}=R_{n,1}=10^{-8}$; the orange trajectory passes through the point $R_{n,0}=R_{n,1}=10^{-16}$. The inset clearly shows that the orange curve is delayed as compared to the blue one.}\label{Fig:Fig6}
\end{figure}

\begin{figure}
\centering
    \includegraphics[width=0.48\textwidth]{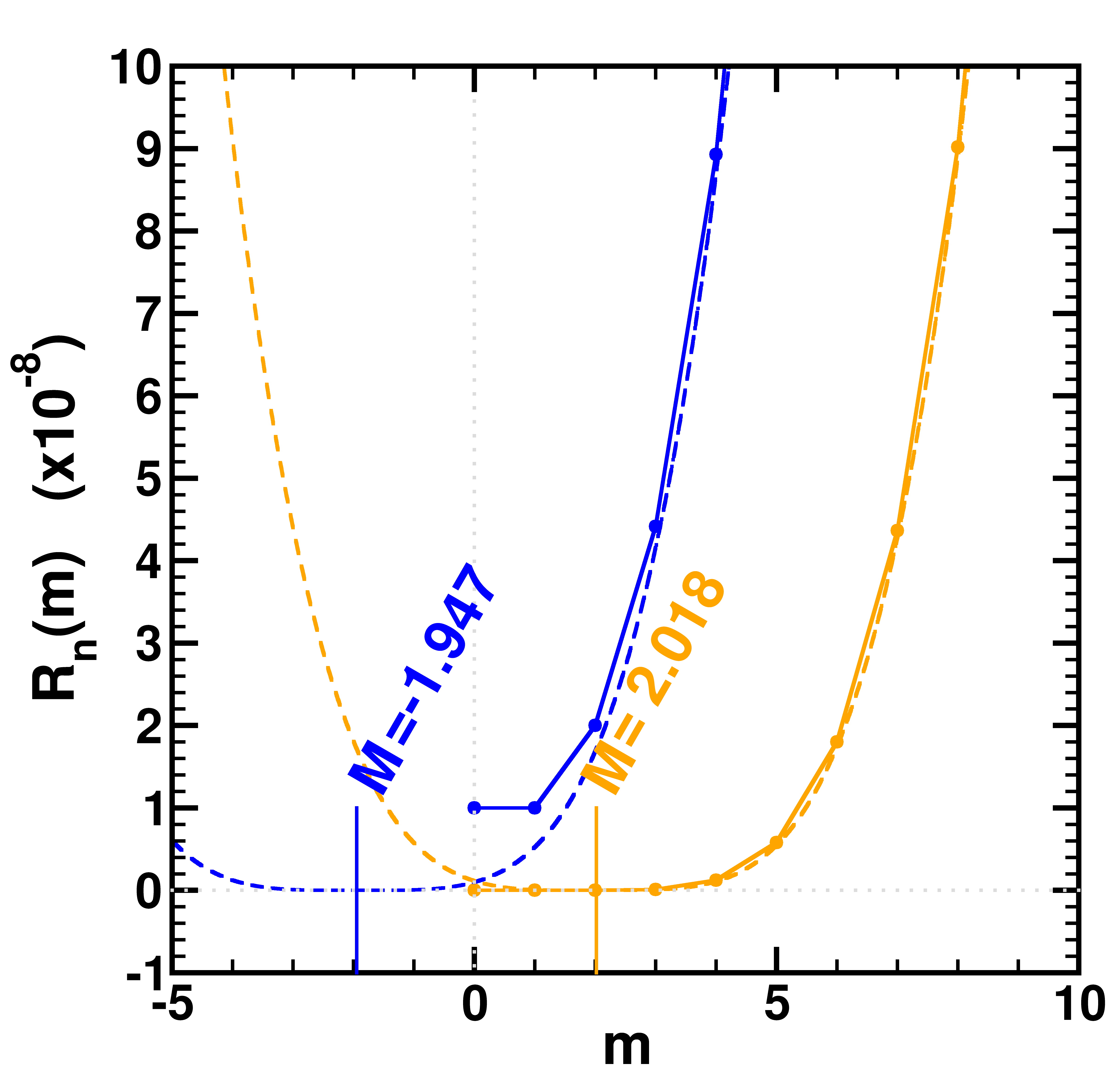}
  \caption{Variables and colors as in Fig.~(\ref{Fig:Fig6}), now shown for small $m$. The dashed curves represent the continuous curves $1/144 (\frac{m-M}{n})^{4}$ that are fitted to the sequences.}\label{Fig:Fig7}
\end{figure}

In Figs.~\ref{Fig:Fig5}--\ref{Fig:Fig7}, we compare initial conditions $R_{n,0}=R_{n,1}=n^{-\frac{2}{(1-a)}}=10^{-8}$ (blue curves) with initial condition $R_{n,0}=R_{n,1}=10^{-8} \times n^{-\frac{2}{(1-a)}}=10^{-16}$ (orange curves) (both with $n=100$ and $a=1/2$).
Figure \ref{Fig:Fig5} shows pairs of $(R_n(m-1),R_n(m))$. These discrete curves can be thought of as parameterized by time: subsequent data points are a temporal distance of $1/n$ apart.
Figure \ref{Fig:Fig6} shows the same two sequences $R_n(m)$ as a function of $m (=nt)$, for large $m$: at this scale, the sequences on the one hand and the continuous curves on the other hand nearly coincide, allowing an excellent fit between them. We see that the orange curve reaches the distance of, \textit{e.g.}, 50 at larger $m$ (i.e., later in time) than the blue curve. Therefore, the orange curve is delayed compared to the blue one, which is consistent with the curves' $M$-values.
Figure \ref{Fig:Fig7} shows the two sequences $R_n$ as a function of $m$, now for small $m$: at this scale, the sequences and the continuous curves are qualitatively different, although the fit between them is excellent for large $m$, as we saw in fig.~\ref{Fig:Fig6}. The fitted values of $M=nT$ do not correspond to the minimum of the sequences, which occurs at $m=0$ for both.

\subsubsection{Results for a general case}
We need to vary both initial conditions $R_{n,0}$ and $R_{n,1}$ independently to study the dependence of the discrete perturbation on $T$. In this section, we study this dependence systematically; therefore, we no longer require the initial grid velocity to be zero.

\begin{figure*}
\centering
  \includegraphics[width=0.9\textwidth]{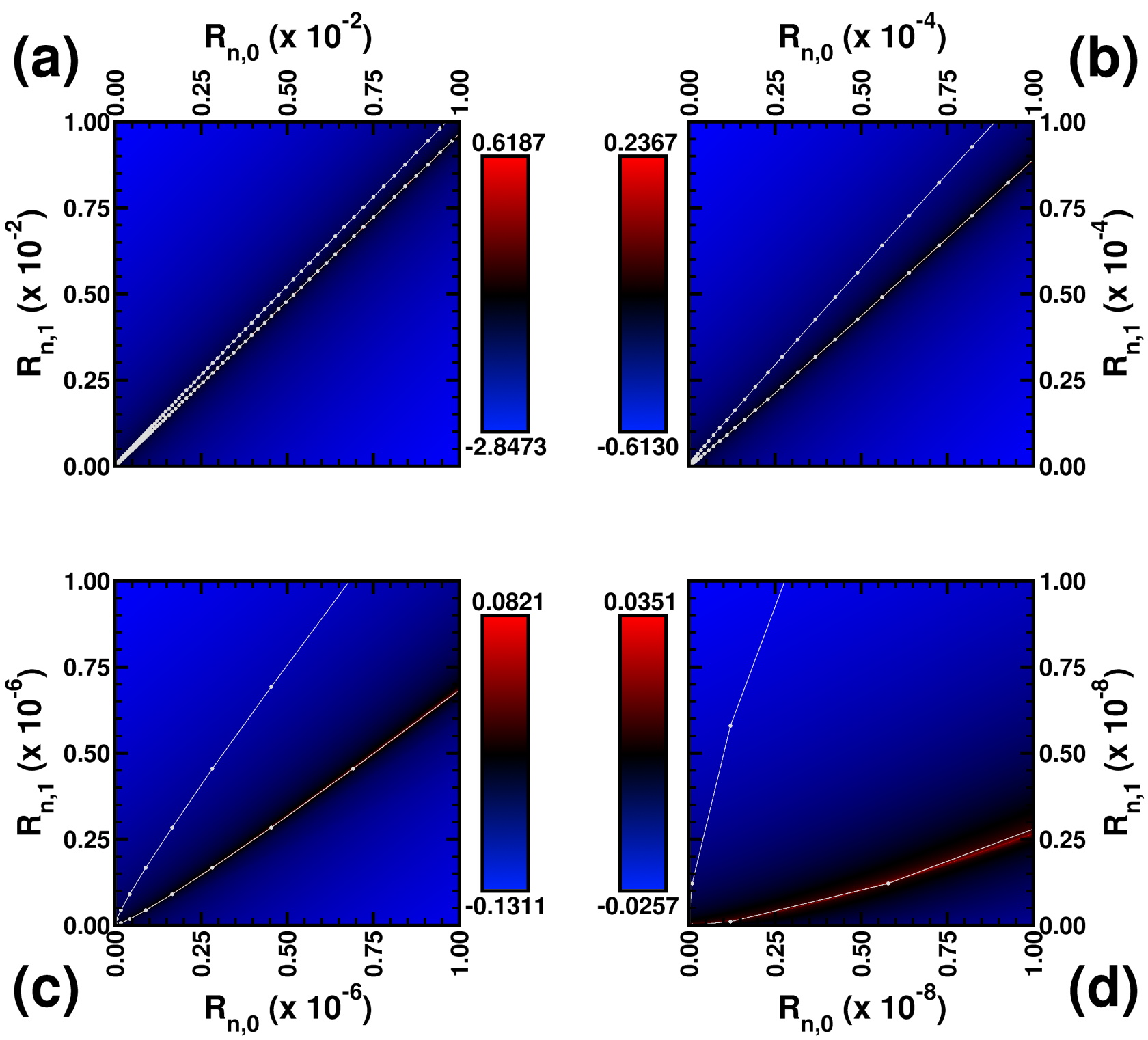}\\
  \caption{Values for $T$, the parameter from the continuous solution that corresponds to the onset of the movement, represented by the color scale in the interval indicated besides each panel. Values for $T$ are obtained numerically from a fit to $R_n(m) = \frac{1}{n^2} R_n(m-1)^a + 2 R_n(m-1) - R_n(m-2)$ with $a=1/2$ in the interval $R_{n,0} \in [0, \frac{1}{n^m} ]$ (horizontal) and $R_{n,1} \in [0, \frac{1}{n^m} ]$ (vertical) with $n=100$ and $m=1$ [panel (a)], $m=2$ [panel (b)], $m=3$ [panel (c)], and $m=4$ [panel (d)]. The overlay shows in gray one trajectory passing through $(R_n(m-1),R_n(m-2))=(10^{-12}, 10^{-12})$.}\label{Fig:Fig9}
\end{figure*}

We wrote a program in visual Pascal (Delphi), which allows us to study the effect of initial conditions in the recurrence equation (\ref{Eq:FiniteGridCauchy}) on the fitted $T$-value understood as $M/n$ via (\ref{Mfit}).\cite{OurCode} We use our program to determine the $T$-values and to represent them using a color scale: see fig.~\ref{Fig:Fig9} for an example of the output. The legend in the figures indicates the range of the $T$-values.
In practice, the $R_{n,0}$ and $R_{n,1}$ intervals start at a number slightly higher than zero (and much smaller than the upper bound): otherwise the singular solution (at $R_{n,0}=0, R_{n,1}=0$) is in the field of view% (at the left bottom corner)
, dominating the $T$-scale.

It is instructive to see the results of our program combined with particular sequences or trajectories. This is shown in fig.~\ref{Fig:Fig9}. The part of the trajectory below the main diagonal corresponds with a mass moving toward the apex and coincides with positive $T$-values.

In general, we see that solutions with a positive $T$-value, visible as a narrow red band in the figures, mainly occur for $R_{n,1}$ smaller than $R_{n,0}$ (below the main diagonal $R_{n,0} = R_{n,1}$). This is to be expected: if the (sufficiently small) initial grid velocity is negative (i.e., directed toward the top), the mass first moves toward the apex, before sliding off, thus increasing the (apparent) $T$.
However, for fixed $R_{n,0}$, $R_{n,1}$ cannot be chosen arbitrarily small; otherwise, we select a trajectory that goes over the top and slides off on the other side, resulting in a smaller positive or negative $T$.
For positive grid velocities, the mass slides off the mound monotonically with a smaller or more negative apparent $T$ at large $m$ as compared to the same $R_{n,0}$ with $V_{n,0}=0$.

\begin{figure}
\centering
  \includegraphics[width=0.48\textwidth]{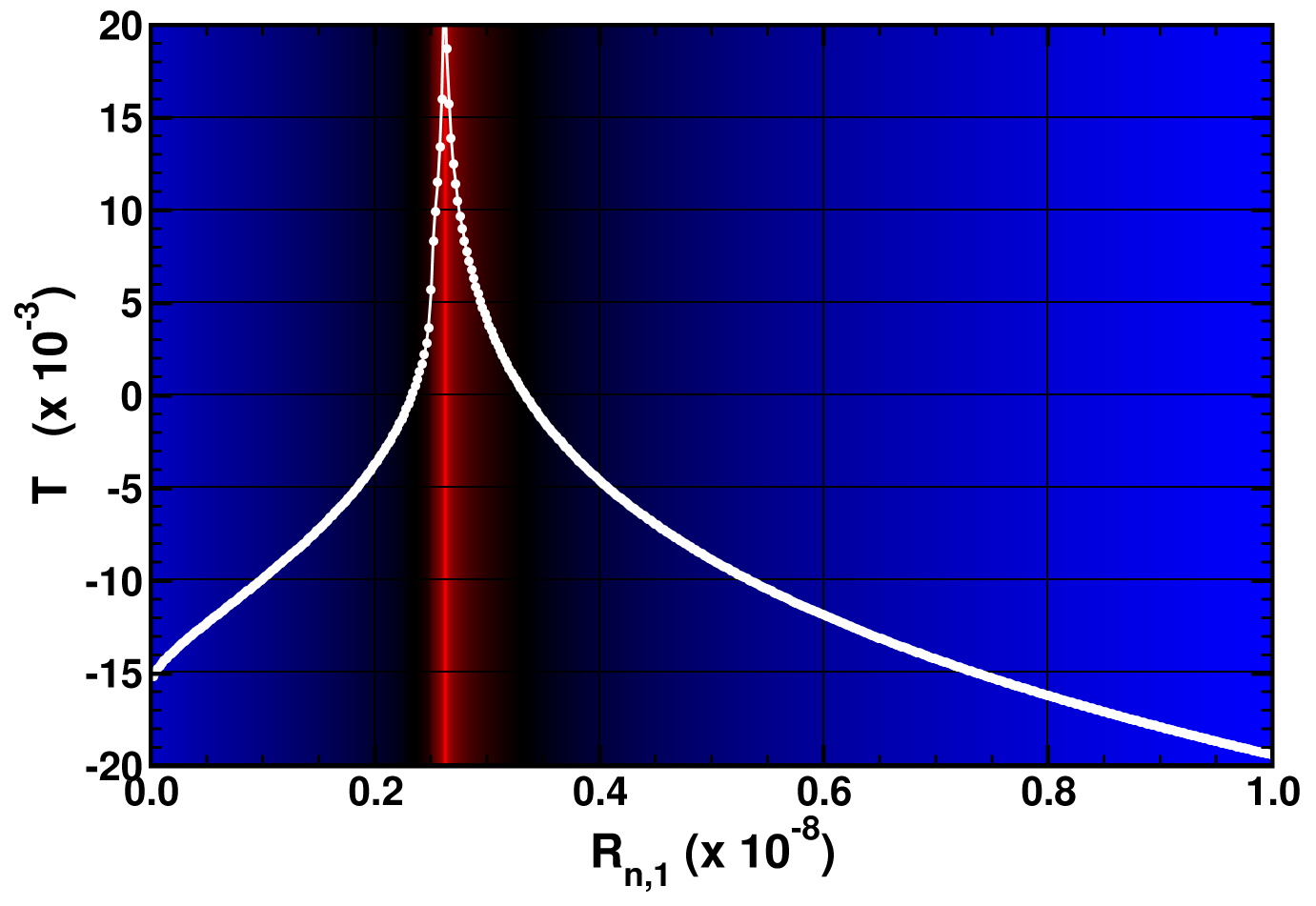}\\
  \caption{Dependence of $T$ on $R_{n,1}$ at constant $R_{n,0} = n^{-\frac{2}{(1-a)}}$, which is $10^{-8}$ for $a=1/2$ and $n = 100$.}\label{Fig:Fig10}
\end{figure}

We also studied the dependence of $T$ on $R_{n,1}$ (keeping $R_{n,0}$ fixed). As an example, we considered $a=1/2$, $n = 100$ and $R_{n,0} = n^{-\frac{2}{(1-a)}} = 10^{-8}$; this corresponds to the right-hand edge in Fig.~\ref{Fig:Fig9}.d.
When $R_{n,1}$ is varied from 0 to $R_{n,0}$, $T$ monotonically increases to an asymptote (located near $R_{n,1} = 0.262~162~16 \times 10^{-8}$) and then monotonically decreases. This is shown in fig~\ref{Fig:Fig10}. (Since we do not have an analytic solution to the recurrence equation, we cannot determine the position of the asymptote analytically either.)

Continuing with the example, the initial condition $R_{n,1}=R_{n,0}$ corresponds to a mass that is released from an arc distance of $10^{-8}$ with grid velocity zero. As we already discussed in Sec.~\ref{Sec:resultsVn0is0}, it immediately starts sliding off from the initial side. This leads to a negative $T$-value.
Initial conditions with $R_{n,1}$ between the asymptote and $R_{n,0}$ correspond to a mass that is released from the same distance, but now with a positive velocity toward the top. As $R_{n,1}$ is decreased in this interval, the mass moves closer to the apex before sliding down, leading to a monotonic increase in the $T$-value toward the asymptote.

The initial condition $R_{n,1}=0$ corresponds to a mass that is released from a distance of the apex with a velocity directed toward the apex, such that the mass already reaches the top at $m=1$ (or $t=\frac{1}{n}$): this leads to a mass that rapidly slides off the dome at the other side, also corresponding with a negative $T$-value.
As $R_{n,1}$ is increased  up to the aforementioned asymptote, the velocity decreases, leading to a slower descent from the other side, hence the monotonically increasing $T$-values.

The slope of the $T$-curve in the $R_{n,1}$-interval between 0 and the asymptote is characterized by an infinite sequence of points of inflection. The first one occurs at $R_{n,1}=0$, the second one at $R_n(2)=0$, the third one at $R_n(3)=0$, \textit{etc}. In general, $R_n(2)= \frac{1}{n^2} R_{n,1}^a + 2R_{n,1} - R_{n,0}$. Solving for $R_{n,1}$ in the case where $R_{n,0} = 10^{-8}$ and $R_n(2)=0$ yields $R_{n,1}= 0.25 \times 10^{-8}$ exactly. In principle, the position of the other points of inflection can be computed from the general expression for $R_n(m)$, but this becomes impractical quickly. [For instance, the relevant equation for the inflection point corresponding with $R_n(3)=0$ is $R_n(3) = \frac{1}{n^2} (\frac{1}{n^2} R_{n,1}^a + 2 R_{n,1} - R_{n,0})^a + 2 \frac{1}{n^2} R_{n,1}^a + 3 R_{n,1} - 2 R_{n,0}$.] Instead of the analytic approach, we have determined numerically that the third and fourth inflection points occur around $R_{n,1}= 0.262~071 \times 10^{-8}$ and $R_{n,1}= 0.262~162~153~8 \times 10^{-8}$, respectively. The position of the asymptote can be regarded as the limit of this sequence of inflection point positions, but this does not yield a more practical way of computing it.

\begin{figure*}
\centering
  \includegraphics[width=0.9\textwidth]{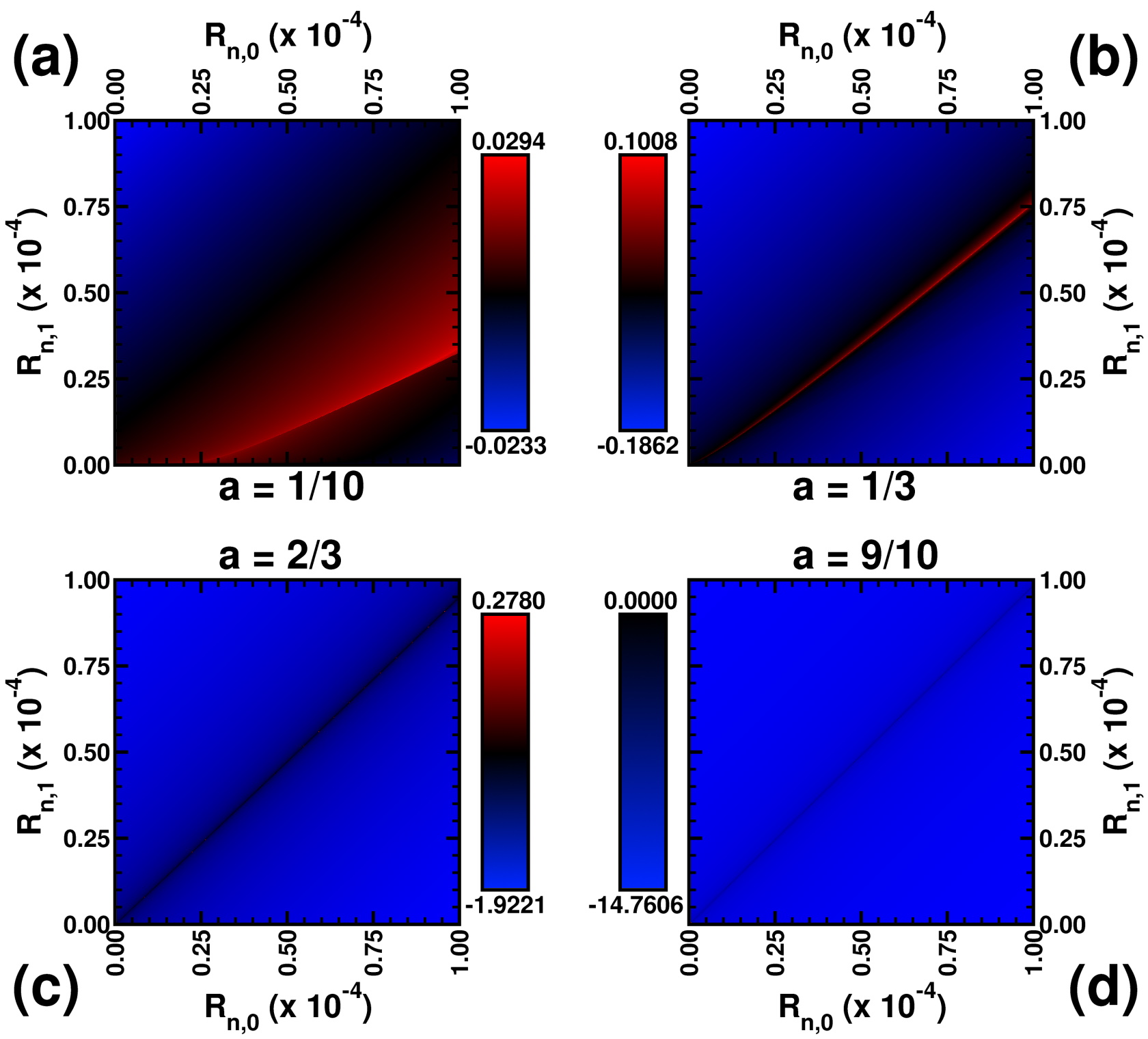}\\
  \caption{Values for $T$, the parameter from the continuous solution that corresponds to the onset of the movement, represented by the color scale in the interval indicated besides each panel. Values of $T$ are obtained numerically from a fit to $R_n(m) = \frac{1}{n^2} R_n(m-1)^a + 2 R_n(m-1) - R_n(m-2)$ with $n=100$ in the interval $R_{n,0} \in [0, 10^4 \times n^{-\frac{2}{(1-a)}}]$ (horizontal) and $R_{n,1} \in [0,  10^4 \times n^{-\frac{2}{(1-a)}}]$. Panel (a): $a=1/10$. Panel (b): $a=1/3$. Panel (c): $a=2/3$. Panel (d): $a=9/10$.}\label{Fig:Fig11}
\end{figure*}

So far, all numerical results we have shown were for Norton's dome ($a=1/2$). Figure \ref{Fig:Fig11} presents results obtained by varying $a$. As $a$ increases, the region with positive $T$-values becomes narrower and its slope approaches the main diagonal. In other words, increasing $a$ looks like zooming out and decreasing $a$ looks like zooming in as compared to the intermediate case where $a=1/2$. Quantitatively, this scaling behavior is consistent with the scale factor, $n^{-\frac{2}{(1-a)}}$.

\subsection{Results in the $\alpha$-limit\label{Sec:ResultsAlphaLimit}}
\begin{figure*}
\centering
  \includegraphics[width=0.9\textwidth]{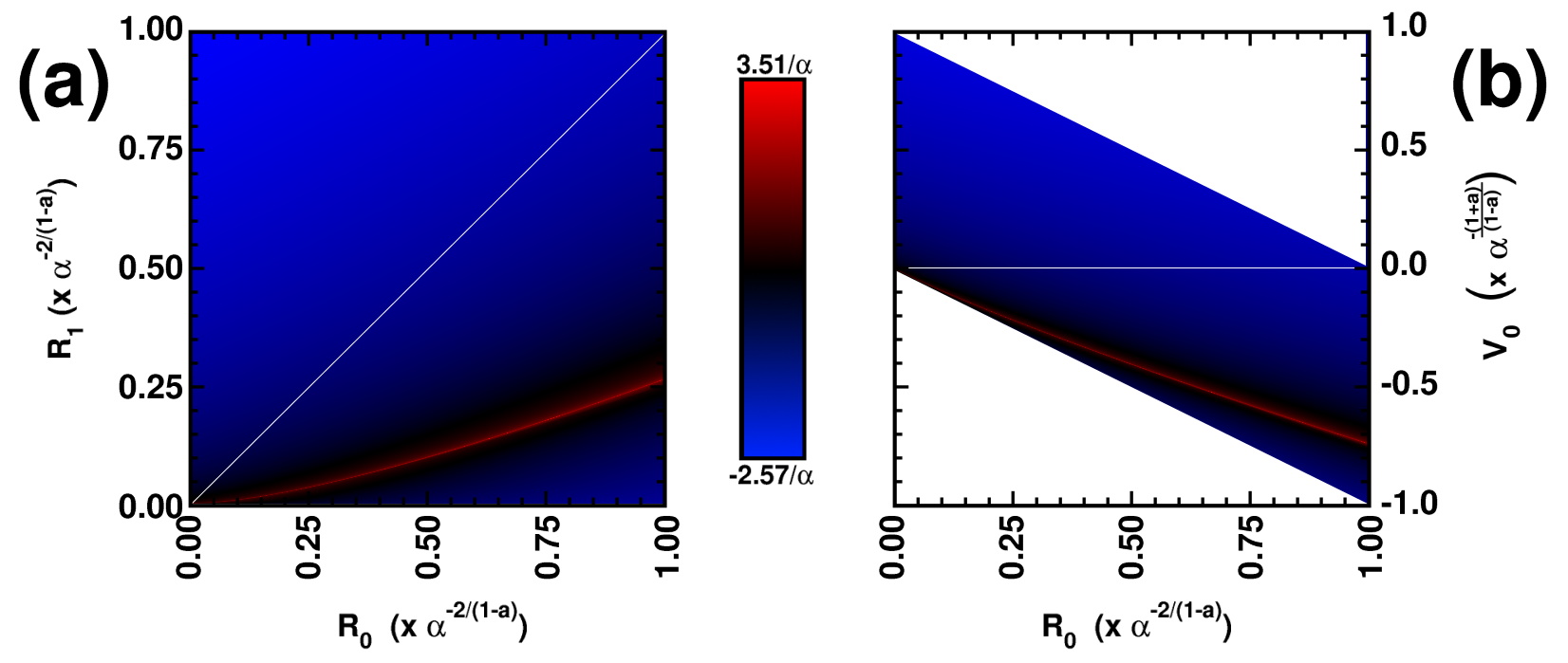}\\
  \caption{Two representations of the same numerical results for $M$, where $M$ is the parameter in $F^a(m) = \left( \frac{(1-a)^2}{2(1+a)} \right)^{\frac{1}{1-a}} \left( \frac{m - M}{\alpha} \right)^{\frac{2}{1-a}}$ [which differs from $R(m)$ by an infinitesimal for infinite $m$], represented by the color scale. 
   Panel (a): $(R_0,R_1)$-plane. Panel (b): $(R_0,V_0)$-plane. The diagonal in panel (a) corresponds to $R_0=R_1$: that is the horizontal axis in panel (b), i.e., $V_0=0$. Values of $M$ are represented by the color scale in the interval indicated besides each panel.}\label{Fig:Fig14}
\end{figure*}

For our purposes, it is crucial that we do not introduce any finite perturbations (as in Sec.~\ref{Sec:Euler}), but that we keep the standard part of the initial values $R_0$ and $V_0$ in (\ref{Eq:HyperfiniteGridCauchy}) exactly zero. In order to achieve this, we take the $\alpha$-limit of sequences of finite grid Cauchy problems, for which the perturbations become infinitesimal in this limit.
We focus on the subset $\left[0, \alpha^{-\frac{2}{(1-a)}}\right]_{^*\mathbb{R}}$ of both $R_0$ and $R_1$ because we know the scaling behavior of the finite discrete functions on the corresponding sequence of finite intervals $\left[0, n^{-\frac{2}{(1-a)}}\right]_{\mathbb{R}}$. Computationally, the results for $M$ do not change, even when we change $n$ (and $a$), as long as we scale the plots like this. Since the result on this scale is exactly the same for every finite $n$ above a certain threshold, alpha-theory guarantees that this scaling behavior holds in the $\alpha$-limit as well. This way, we can plot panel (a) of fig.~\ref{Fig:Fig14} using standard numerical simulations.

\begin{figure}
\centering
  \includegraphics[width=0.48\textwidth]{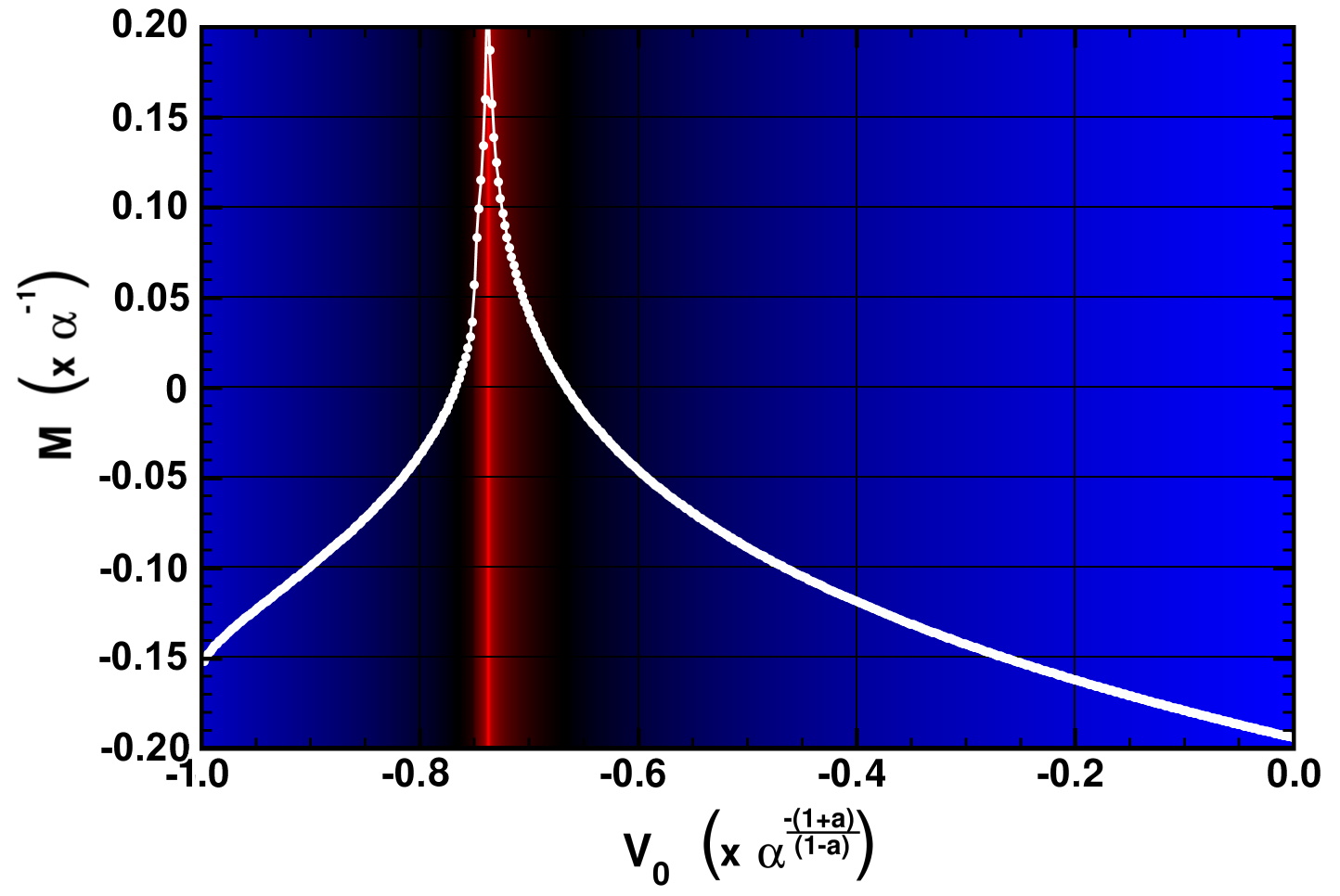}\\
  \caption{Dependence of $M$ on $V_{0}$ at constant $R_{0} = \alpha^{-\frac{2}{(1-a)}}$.}\label{Fig:Fig15}
\end{figure}

For finite $m$, the solution to (\ref{Eq:HyperfiniteGridCauchy}), $R(m)$, is a finite sum of infinitesimals; therefore, its standard part is zero.
For infinite $m$, $R(m)$ is a hyperfinite sum that in general need not be infinitesimal, even when all the terms are.
For example, for the sequence of finite difference equations with initial conditions $R_{n,0} = R_{n,1} = n^{-\frac{2}{(1-a)}}$, the difference between the sequence $R(m)$ and $F^a_n(m) = \left( \frac{(1-a)^2}{2(1+a)} \right)^{\frac{1}{1-a}} \left( \frac{m - M}{n} \right)^{\frac{2}{1-a}}$ decreases faster than $1/m$, where we find numerically that $M$ is about equal to $-1.947~417$.
Therefore, when we take the $\alpha$-limit of this sequence of finite grid Cauchy problems, with $R_0 = R_1 = \alpha^{-\frac{2}{(1-a)}}$, we find the hyperfinite grid solution:
$R(m) = \left( \frac{(1-a)^2}{2(1+a)} \right)^{\frac{1}{1-a}} \left( \frac{m-(-1.947~417)}{\alpha} \right)^{\frac{2}{1-a}}$. For values of $m$ such that $\frac{m}{\alpha}$ is finite, we take $t=st(\frac{m}{\alpha})$. Then, $st(R(m))$ corresponds with the general solution (\ref{Eq:MalamentSolution}) with $T=st(-\frac{-1.947~417}{\alpha})=0$.
Therefore, for all $m$, $st(R(m)) = r(t)$, where the latter is one of the solutions to the continuous Cauchy problem. In particular, we find the undelayed solution: $r(t)= \left( \frac{(1-a)^2}{2(1+a)} \right)^{\frac{1}{1-a}} \left( t \right)^{\frac{2}{1-a}}$.
As we see in fig.~\ref{Fig:Fig15}, to obtain a standard solution with a non-zero $T$, we have to pick a $V_{0}$-value that is finely tuned to $R_{0}$. This observation is directly relevant for the assignment of probabilities in Sec.~\ref{Sec:Probability}.

As explained in Sec.~\ref{Sec:AlphaTheory}, the indeterminism in the standard model (continuous, without infinitesimals) can be interpreted as being due to rounding off the infinitesimals from the hyperfinite model. We have now seen an example of this in terms of our toy model.
Finally, this attribution or correspondence allows us to assign probabilities to the standard solutions.

\section{Using the hyperfinite grid differential equation to assign probabilities to the standard solutions\label{Sec:Probability}}
Faced with the family of regular solutions (\ref{Eq:MalamentSolution}) to (\ref{Eq:MalamentMounds}), it might be tempting to impose a probability density on $T$ directly: a uniform probability measure on $T$ perhaps? This approach faces two problems. The first one is an immediate consequence of the fact that $T$ may be arbitrarily large: there is no standard countably additive probability measure that is uniform over an infinite support. This problem may be overcome by adopting a merely finitely additive probability function.
The second problem is that the measure is not robust under reparameterizations; therefore, one needs an argument to favor a uniform measure over $T$, rather than over $1/T$, $\log(T)$, or some other transformation.

Instead of imposing a probability measure on $T$ directly, for which we know no principled way of choosing one, we approach the issue differently. Starting from the hyperfinite model, we first consider hyperreal initial conditions $(R_0, V_0)$ that are randomly chosen from a suitable interval that guarantees that they are both infinitesimal and then compute the resulting probabilities for obtaining the singular solution and for the values of $T=st(M/\alpha)$ associated with the family of regular solutions.
The first step amounts to assuming a uniform prior on the phase space. Now, we only face the second problem: arbitrary coordinate transformations lead to infinitely many representations of the same phase space. Our next question, then, is how to make a principled choice here.

\subsection{Phase space for uniform, random sampling}
In (\ref{Eq:MalamentMounds}) and (\ref{Eq:HyperfiniteGridCauchy}), we have used the Newtonian formalism with two generalized coordinates: the arc length and the arc (grid) velocity. It is well known that measures on the phase space change due to coordinate transformations. To report the results of our numerical experiments, we have used the arc length at two different times as the phase space ($R_0$ and $R_1$), rather than the initial arc length and the grid velocity ($R_0$ and $V_0$), which can be viewed as an example of such a transformation.

To make a principled choice, we take our cue from statistical mechanics. In Ref.~\citenum{Goldstein:2012}, Goldstein reviewed arguments (going back to Boltzmann) to the effect that a non-arbitrary choice for the probability measure is a uniform measure that is invariant under the dynamics.
Statistical physicists had to resolve this issue because they often appeal to the notion of typicality in the sense of `almost all' trajectories. Clearly, such statements only make sense relative to a specific measure, which turns out be the Lebesgue measure on the phase space. This choice is motivated by Liouville's theorem, which implies that the Lebesgue measure on the phase space conserves probability.
Outside of statistical mechanics, the approach to associate a unique probability measure to a random selection of initial conditions via invariance requirements is also well-known from the work of Jaynes,\cite{Jaynes:1973} which led to many applications of such maximum information entropy (MaxEnt) methods.

In other words, we need coordinates such that the density of states is conserved on the phase space. For this, we have to consider the Lebesgue measure on the canonical coordinates from Hamiltonian mechanics (such that Liouville's theorem applies). For our problem, the canonical coordinates are the arc length and the conjugate momentum, which equals mass times the arc velocity. Since we assumed a unit mass, the phase space spanned by $(R_0, V_0)$ is the proper starting point for a probabilistic analysis that allows us to start from a uniform probability distribution (MaxEnt).
Hence, we use $(R_0, V_0)$ to represent the initial conditions [where $V_0 = \alpha (R_1-R_0)$], as shown in panel (b) of Fig.~\ref{Fig:Fig14}.

\begin{figure}
\centering
  \includegraphics[width=0.48\textwidth]{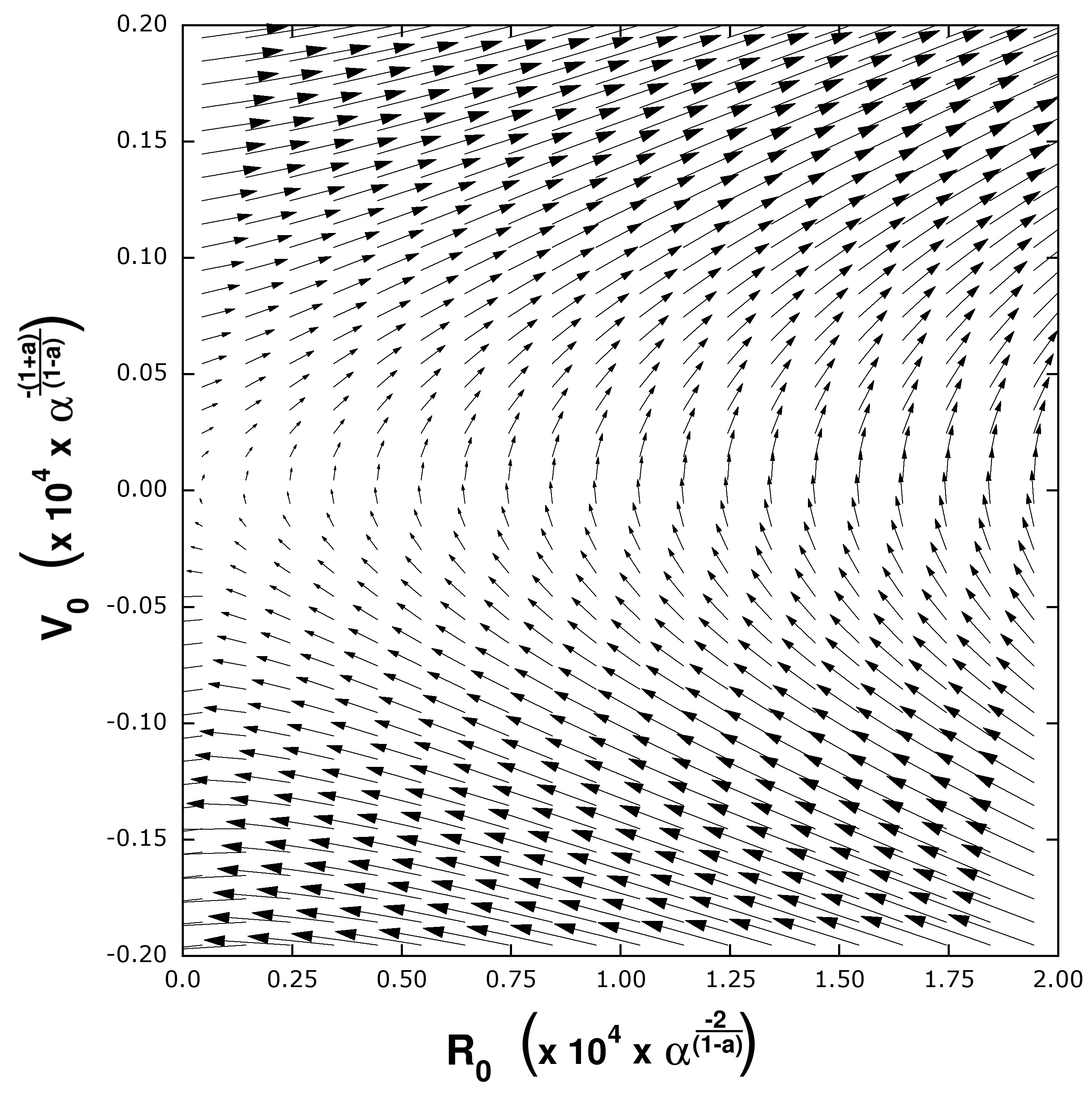}\\
  \caption{Numerical results for the vector field on the phase space $(R_0,V_0)$.}\label{Fig:Fig13}
\end{figure}

Therefore, our prior probability is motivated by the dynamics, which privileges the uniform measure on the $(R_0, V_0)$-plane. This is what we take selecting `random' initial conditions to mean and how to reason about `typical' results. However, this is a defeasible choice: if there is any background information on how the initial position and velocity are realized (due to preparation or post-selection of the system), we should adapt the prior in light of it. For instance, one might consider a process that aims to place the mass as close to the top as possible with a velocity as close to zero as possible and then selects systems of which the real-valued initial position and velocity are indeed zero. In such a case, it is clear that the prior does not come from the dynamics of the system under study, but from an independent placement mechanism. In the case of such an external influence, the Cauchy problem does not contain all the information about the system; therefore, it does not suffice to determine the probabilities.
For this example, we may want to consider a Gaussian distribution around the origin in the $(R_0, V_0)$-plane instead of a uniform distribution.
Fortunately, as we will see in Sec.~\ref{Sec:Counting}, our main results (in terms of probabilities for the $T$-values) turn out to be quite robust: they are valid for a uniform measure over $(R_0, V_0)$ as well as any Gaussian or other finite transformations of it.

The vector field on this phase plane is shown in Fig.~\ref{Fig:Fig13}, where the symmetry between the upper and lower half of the plane shows that the dynamics is time reversible. If we reinterpret $R$ as a position rather than an arc length and include the opposite side of the slope as negative $R$-values (not shown), the vector field shows the signature of a saddle point at the singularity $(0,0)$ (indicating an unstable equilibrium).

\subsection{From measuring initial conditions to probabilities\label{Sec:Counting}}
We now propose to measure the probability of sets of standard solutions as the standard part of the normalized area of the set of corresponding initial values, $(R_0,V_0)$, in the hyperfinite grid model.
Since there is no such thing as the `largest infinitesimal', we have to normalize on $R_0$ and $V_0$ being in a particular interval of infinitesimals. For this purpose, we select $R_0 \in [0,\frac{1}{\alpha}]$ and $V_0 \in [-\frac{1}{\alpha},\frac{1}{\alpha}]$; therefore, the normalization factor is $\frac{2}{\alpha^2}$.
We can now represent any event as a subset of $[0,\frac{1}{\alpha}] \times [-\frac{1}{\alpha},\frac{1}{\alpha}]$ and consider its probability as the standard part of the area of the set divided by $\frac{2}{\alpha^2}$.
(This approach is similar to that of Ref.~\citenum{Benci_etal:2010}, where it is connected to a Loeb measure.\cite{Loeb:1975})

All the events we have discussed in Sec.~\ref{Sec:ResultsAlphaLimit} were contained in $(R_0,V_0) \in \left[0, \alpha^{-\frac{2}{(1-a)}}\right] \times \left[-\alpha^{-\frac{(1+a)}{(1-a)}},\alpha^{-\frac{(1+a)}{(1-a)}}\right]$, which is a strict subset of the proposed reference class (for every $a$) and has an infinitesimal normalized area of $\alpha^{-\frac{(1+3a)}{(1-a)}}$.

We already observed that exactly one combination of hyperreal-valued initial conditions leads to the equilibrium solution: $R_0=0$ and $V_0=0$. This means that the singular solution has zero area. Although it is not logically impossible for it to happen, it does not happen almost surely. This assignment is very robust: it holds not only for the uniform prior on $(R_0,V_0)$, but for any prior that does not assign more than an infinitesimal portion to the singleton $(0,0)$.
All other initial conditions, in $[0,\frac{1}{\alpha}] \times [-\frac{1}{\alpha},\frac{1}{\alpha}] \setminus \{(0,0)\}$, are associated with a regular solution. They carry unit probability; therefore, a regular solution happens almost surely. This settles our first research question.

Our second research question asked how to assign relative probabilities to the $T$-values in the family of regular solutions. The key to answer this lies in our observation that the relation between the infinitesimal initial conditions and the $T$-parameter in the corresponding continuous solution is strongly non-linear [where $T=st(M/\alpha)$].
Figure \ref{Fig:Fig15} shows that, for the specific choice where $R_0=\alpha^{-\frac{2}{(1-a)}}$, the interval where $M$ is positive covers a non-infinitesimal fraction of $V_0 \in \left[ -\alpha^{-\frac{(1+a)}{(1-a)}}, 0 \right]$ (about 10\%), but this interval itself is only an infinitesimal fraction of the entire $V_0$-range, $[-\frac{1}{\alpha},\frac{1}{\alpha}]$.
Therefore, the normalized area corresponding to positive $M$-values is infinitesimal.
Moreover, almost all positive $M$-values are such that $M/\alpha$ is infinitesimal: since $M$ increases so fast in the region where it is positive, the interval of $V_0$-values that correspond to a non-infinitesimal value is infinitesimal compared to the interval where $M$ is positive (fig.~\ref{Fig:Fig15}).
(Due to this highly non-linear dependence of $T$ on $V_0$, for fixed $R_0$, it was also difficult in the numerical experiments with the corresponding finite difference equations to find explicit examples of large, positive $T$-values.)
The normalized area corresponding to negative $M$-values is unity minus an infinitesimal. The negative $M/\alpha$-values are all infinitesimal.
Therefore, for $R_0=\alpha^{-\frac{2}{(1-a)}}$, we find the undelayed standard solution [i.e., $T=st(M/\alpha)=0$] almost surely.

These observations hold in general, across all $R_0$: almost all infinitesimal initial conditions correspond to the undelayed standard solution and only an infinitesimal proportion of all infinitesimal initial conditions correspond to standard solutions with $T=st(M/\alpha)>0$.
Hence, if we assume a uniform distribution of the infinitesimal initial conditions, we arrive at the following probabilities for the standard solutions:
\begin{itemize}
  \item The probability of the mass staying at the apex of any of Malament's mounds indefinitely (singular solution) is zero.
  \item The normalized area of initial conditions in the hyperfinite grid model corresponding to the mass staying at the apex of any of Malament's mounds for some observable time is infinitesimal; therefore, the probability is zero.
  \item The normalized area of initial conditions in the hyperfinite grid model corresponding to the mass immediately starting to slide off the apex of Norton's dome or any of Malament's mounds is one minus an infinitesimal; therefore, the probability is unity.
\end{itemize}

In conclusion, a point mass with velocity zero at the apex of any frictionless Malament's mound in a uniform gravitational field will immediately start sliding off the dome almost surely.
If the initial conditions are external to the Cauchy problem, other priors can be considered and our methodology may yield another result.
For instance, for an asymmetrical measure on the $(R_0,V_0)$-plane, the conclusion will be different.
However, the above conclusion continues to hold if the uniform measure on the $(R_0,V_0)$-plane is replaced by a symmetric Gaussian (cut-off for non-infinitesimal values, since they contradict the conditions set by the standard initial value problem).

In addition, for a mass sliding toward the apex (from a finite distance and with finite velocity), reaching it with a standard velocity of exactly zero, it will either slide off from the opposite side immediately or slide back immediately (depending on the precise infinitesimal position and velocity values), almost surely.
In this case, the initial values are sampled from a different part of the phase plane, but it remains the case that those corresponding to any measurable delay are of infinitesimal measure as compared to those yielding no measurable delay.

So far, we have only presented equations and results that rely on distances to the apex. By adding sign information, we can keep track of which side of the two-dimensional cross section of the dome the mass is on. We find that the probability of sliding off on either side of the dome is $1/2$. This can be seen directly from the symmetry of the extended phase plane (not shown) in combination with the uniform prior. Moreover, observing the initial infinitesimal position and velocity allows us to predict the final descent direction. Therefore, unlike the original model, the hyperfinite model does not exhibit spontaneous symmetry breaking: either the symmetry is broken at $m=0$ or the solution remains symmetric (when $R_0=0$ and $V_0=0)$.
Hence, the symmetry breaking in the standard model may be thought of as being due to rounding off the infinitesimals in the hyperfinite model.

\section{Discussion and Conclusion\label{Sec:Disc}}
First, we comment on possible applications of this work. Then, we draw general conclusions.

\subsection{Relation to contemporary hydrodynamics literature\label{Sec:Hydrodyn}}
In our paper, we focused on a toy example. However, differential equations with a non-Lipschitz singularity are prevalent in the context of physical applications, such as shock formation and turbulent flows; a case that is widely studied is that of the Burgers equation (a first-order, non-linear partial differential equation) in the inviscid limit (i.e., the limit of viscosity to zero, which is equivalent to the infinite limit of the Reynolds number). In Sec.~\ref{sec:Intro}, we already mentioned some publications that used probabilistic approaches to such problems. Here, we briefly review results from this literature. Where possible, we connect their results to ours. Fully comparing, `translating', and contrasting the cited works to our methodology, however, would warrant a separate study, much more extensive than the current paper. In any case, our study shows that it will be crucial to pay due attention to the order of the (standard) limits: when the limit of finite perturbations to zero is taken as the first step, there is no way to recover the probabilities associated with various rates of convergence afterward. Alpha-limits have the advantage of retaining this asymptotic behavior automatically by encoding it into distinct infinitesimals.

In the context of passive scalar transport in a turbulent velocity field, E and Vanden-Eijnden\cite{EVandenEijnden:2000} introduced `generalized flows', which are families of probability distributions on the space of solutions to a non-Lipschitz ODE. They started from a first-order ODE for a non-Lipschitz velocity field and considered probability distributions on the set of solutions: either as a probability measure on the path-space or as transition probabilities (which degenerates to unit probability mass at the unique path in the case of Lipschitz continuity). They considered the analogy to stochastic ODEs with a random (white noise) velocity field and also two natural regularizations of the problem, which do not always give identical results.

Building on this, E and Vanden-Eijnden\cite{EVandenEijnden:2003} presented some examples, including a first-order analogous case to the second-order ODE that we discussed. Applied to our case, their approach relies on a stochastic process to determine the time and the initial direction of the descent from the top. Like in our approach, the singular solution has measure zero and both directions of descent are equiprobable. They considered transition probability distributions to characterize the random field, which they used to define a generalized flow for the non-Lipschitz ODE.

Falkovich \textit{et al.}\cite{Falkovich_etal:2001} compared chaotic (deterministic) behavior with exponential separation and truly turbulent (stochastic) behavior with explosive separation (with power law scaling): only in the second case do infinitesimally close trajectories separate in finite time. In the inviscid limit, the ODE becomes non-Lipschitz, allowing for multiple Lagrangian trajectories. They considered a statistical description of the trajectories, in terms of a stochastic Lagrangian flow, which allows, for instance, the study of averages.

E and Vanden-Eijnden\cite{EVandenEijnden:2000,EVandenEijnden:2003} showed that different regularization processes give rise to different generalized flows, without one of them being uniquely well-motivated by the underlying physical context. In more recent work, however, Mailybaev \textit{et al.}\ did propose a way to assign a unique (i.e., independent of the regularization method) statistical probability distribution to the Burgers equation in the inviscid limit,\cite{Mailybaev:2016} as well as a class of ODEs that also become non-Lipschitz in this limit.\cite{Drivas_etal:unpublished}

After a trajectory encounters such a singularity in finite time (known as `blowup'), the evolution is no longer deterministic: there are continuum many solutions.
One way to understand this is that the initial state represents a Dirac-delta distribution of initial conditions: whereas this remains a Dirac-delta distribution for fully deterministic evolutions, non-Lipschitz singularities make the delta-distribution spread out to a `spontaneous' probability distribution---a phenomenon known as `spontaneous stochasticity'. This phenomenon may also have a quantum-mechanical analog. \cite{EyinkDrivas:unpublished}

Using methods related to those in our paper, we can re-interpret the Dirac-delta distribution as a function from non-standard analysis (an infinitely small Gaussian), non-singular points as finite dispersion (such that infinitesimal differences remain infinitesimal), and the singular point as a place where there is infinite dispersion (such that infinitesimal differences become non-infinitesimal). Viewed as such, the non-Lipschitz indeterminism of the continuous model can be connected to a case of deterministic chaos in a corresponding hyperfinite model.

\subsection{Conclusion}
In this paper, we presented a method for assigning probabilities to the solutions of initial value problems that lack Lipschitz continuity.
First, we linked the differential equation to sequences of finite grid differential equations, which are deterministic and allow for systematic numerical studies.
Second, to avoid introducing any non-infinitesimal perturbations, we considered the $\alpha$-limit of the sequences and their solutions as hyperfinite grid functions.
Third, starting from a uniform prior on the phase space spanned by the canonical coordinates, we assigned probabilities to the standard part of these hyperfinite grid functions, which equal the solutions of the corresponding, continuous Cauchy problem.
Although we set out to find a probability distribution over the solutions, we found unit probability for one single solution (the non-delayed, regular solution). Therefore, although we did not assume uniqueness at the outset, we do find ourselves in agreement with authors who set out to find a unique continuation beyond the non-Lipschitz discontinuity.

Our methodology and results are analogous to the study of fully deterministic chaotic systems without any singularities: in classical chaos, when two systems have initial conditions that differ by a non-infinitesimal amount below the measurement precision, they cannot be distinguished empirically at the start. Yet, the resulting trajectories measurably diverge at some point and these later states can be used to determine their initial positions beyond measurement precision at the time. (However, see Ref.~\citenum{Gisin:2019}, which suggested to reinterpret this as indeterminism after all.)
Likewise, in the case of indeterministic Cauchy problems, the changes at later times can be attributed to infinitesimal differences present at the start in the corresponding hyperfinite model. (These infinitesimal differences are not measurable on any real-valued measurement precision.)
The hyperfinite model produces asymmetric results only when the initial conditions are asymmetric; therefore, we find that the symmetry breaking in the standard model can be thought of as the result of rounding off the infinitesimals.

The current paper focused on toy examples (Malament's mounds), but our methodology is applicable to study other initial value problems that lack Lipschitz continuity. Hence, we hope our approach will be fruitful for application to the study of singularities in hydrodynamics and other blowup phenomena.

%%%%%%%%%%%%%%%%%%%%%%%%%%%%%%%%%%%%%%

\begin{acknowledgments}
We are grateful to an anonymous referee for a very instructive report that greatly helped us to improve the presentation of our methodology, to Vieri Benci for helpful discussions on this topic, and to Christian Maes for feedback on an earlier version.
Part of SW's work was supported by the Research Foundation Flanders (Fonds Wetenschappelijk Onderzoek, FWO) (Grant No. G066918N).
\end{acknowledgments}

\section*{Author Declarations}
\subsection*{Conflict of Interest}
The authors have no conflicts to disclose.

\section*{Data Availability Statement}

The program binary as well as the source code for the numerical study are available in GitHub at \href{https://github.com/DannyVanpoucke/NortonDomeExplorer}{https://github.com/DannyVanpoucke/NortonDomeExplorer}, Ref.~\citenum{OurCode}.

\section*{References}
%\nocite{*}
\bibliography{aipsamp}% Produces the bibliography via BibTeX.

\end{document}